\documentclass[aps,prd,superscriptaddress]{revtex4-2}
\usepackage{amsmath,amssymb,amsthm,bm,graphicx,subfigure,xcolor,cases,multirow}
\usepackage[colorlinks=true,linkcolor=blue,citecolor=blue,urlcolor=black]{hyperref}

\begin{document}
\title{Hydrodynamic properties in soliton field theory}

\author{Qian Chen}
\email{qchen@xjtu.edu.cn}
\affiliation{School of Physics, Xi’an Jiaotong University, Xi’an, 710049, China}

\begin{abstract}
The crucial role of hydrodynamic instabilities in soliton field theory is revealed.
We demonstrate that the essential of soliton formation mechanism is the sound mode instability induced by thermodynamic instability.
This instability triggers phase separation, where new thermal phases are generated to produce solitons.
These solitons can be regarded as a coexistence state composed of a matter phase and a vacuum phase, with an interface providing surface tension to maintain dynamical equilibrium.
The phase separation mechanism naturally allows the existence of vacuum bubbles, characterized by a vacuum phase surrounded by a matter phase with negative pressure.
Furthermore, we show that the soliton interface resembles a fluid membrane, whose interface pressure satisfies a Young-Laplace-type relation, resulting in the emergence of the membrane instability induced by surface tension.
In the thin-wall limit, the dispersion relation is analytically derived.
This instability triggers topological transition of the interface, splitting a cylindrical interface into multiple spheres with a smaller total surface area.
Such results highlight the duality between solitons and fluids, providing a field theory description for hydrodynamics with interfaces.

\end{abstract}

\maketitle

\section{Introduction}\label{sec:I}
A scientific discipline is typically divided into multiple subfields, each with its own set of principles and methodologies.
In physics, an effective classification criterion is spatial scale, such as the atomic realm dominated by quantum mechanics, the classical world where Newtonian mechanics applies, and the cosmic environment where general relativity comes into play.
The hierarchical structure of physical laws has brought significant benefits to researchers, allowing them to delve into a specific field without having to master all underlying theories.
Another romance of physics is commonality, manifested in the fact that systems at different scales follow the same fundamental laws.
A quintessential example is fluid dynamics, which originated in the description of the dynamical behavior of classical liquids and has since been continually extended to a series of analogs, such as quark–gluon plasma \cite{Baym:1983amj,Heinz:2013th,Gale:2013da}, black holes \cite{Damour:1978cg,Thorne:1986iy,Parikh:1997ma}, holographic systems \cite{Policastro:2001yc,Baier:2007ix,Bhattacharyya:2007vjd,Kovtun:2004de}, and solitons \cite{Coleman:1985ki,Chen:2024axd}, confirming the cross-scale universality of this theoretical framework \cite{Andersson:2006nr}.

\subsection{Fluid dynamics}
A fluid is a substance that undergoes continuous deformation under the action of arbitrarily small shear stress, and is ubiquitous in nature, such as liquids and gases.
Beneath this seemingly simple form of matter lie rich and fascinating phenomena, such as superfluidity \cite{Kapitza1938} and turbulence \cite{Reynolds1883}, that have long captivated generations of theoretical and experimental scientists.
The physics of fluids, usually called fluid dynamics, is one of the oldest and most successful theories for describing non-equilibrium systems, combining simplicity and universality.
The simplicity lies in the fact that this framework is constructed solely from thermodynamic quantities and transport coefficients, whereas its universality stems from a single fundamental assumption\footnote{The traditional criterion for the applicability of fluid dynamics is the existence of a clear separation between the microscopic scale $L_{mi}$ and the macroscopic scale $L_{ma}$, quantitatively characterized by the Knudsen number $\mathrm{Kn}=L_{mi}/L_{ma}$. When the Knudsen number is sufficiently small $\mathrm{Kn}\ll 1$, microscopic interactions occur at ultrafast rates over extremely short distances, and the fluid remains in a state near local thermodynamic equilibrium at the macroscopic scale. The deviation is treated as a high order correction to the local equilibrium state (the ideal fluid), manifesting as dissipative effects and admitting a series expansion in powers of the Knudsen number. 
	In the relativistic version, the above phenomenological criterion is replaced by the requirement that gradients around local equilibrium configuration are small compared to the temperature of system, giving rise to a theoretical framework formulated as a gradient series. Remarkably, the relaxation of the near-local-equilibrium assumption has facilitated the development of a broader framework of fluid dynamics applicable to far-from-local-equilibrium regimes \cite{Heller:2013fn,Heller:2015dha,Romatschke:2017vte}.} that continuum systems remain close to local thermodynamic equilibrium throughout the dynamics, thereby obviating the need for an explicit treatment of complex microscopic interactions.

Looking back in history, the story can be traced back to the mid-eighteenth century.
Frederic II of Prussia asked Euler how to raise water to the Sanssouci Palace, located on a hill in Potsdam, to supply the fountains there.
Ultimately, the failure of this project led to the inception of fluid dynamics, known as ideal fluid dynamics \cite{Euler1755}.
Considering an ideal fluid characterized solely by velocity $\vec{v}$, pressure $p$, and mass density $\rho$, this framework consists of the continuity equation and the Euler equation:
\begin{subequations}
	\begin{align}
		\left(\partial_{t}+\vec{v}\cdot\nabla\right)\rho&=-\rho\nabla\cdot\vec{v},\label{eq:1.1a}\\
		\left(\partial_{t}+\vec{v}\cdot\nabla\right)\vec{v}&=-\left(\nabla p\right)/\rho.\label{eq:1.1b}
	\end{align}\label{eq:1.1}
\end{subequations}
These two equations encapsulate the fundamental conservation laws of mass and momentum for non-dissipative fluids, forming the core of fluid dynamics.
Despite its elegance, the ideal fluid dynamics framework assumes the absence of internal microscopic interactions and thus fails to capture deviations from local thermodynamic equilibrium, which manifest as dissipative phenomena.
These effects are systematically incorporated into the Euler equation \eqref{eq:1.1b} through the divergence of a tensor $\boldsymbol{\pi}$ of rank two, known as the viscous stress tensor, formulating the framework of dissipative fluid dynamics:
\begin{equation}
	\left(\partial_{t}+\vec{v}\cdot\nabla\right)\vec{v}=-\left(\nabla p+\nabla\cdot\boldsymbol{\pi}\right)/\rho.\label{eq:1.2}
\end{equation}
The applicability of fluid dynamics requires that the dissipative correction $\boldsymbol{\pi}$ admit a series expansion in powers of the Knudsen number $\mathrm{Kn}$, which is treated as a small parameter quantifying the separation between microscopic and macroscopic scales.
At the local equilibrium limit, the corrected equation \eqref{eq:1.2} should degenerate to the Euler equation, implying $\boldsymbol{\pi}(\mathrm{Kn}\rightarrow 0)=0$.
Truncating the series expansion at linear order yields the well-known Navier-Stokes theory \cite{Navier1823,Stokes1845}.
This extended framework has been extensively validated and applied in numerous fields such as engineering, meteorology, and geophysics, providing a reliable description of realistic fluid behavior.

The turning point of the story came in the mid-twentieth century, when the development of high-energy physics necessitated the unification of fluid dynamics with special relativity, formulating a Lorentz-invariant framework known as relativistic fluid dynamics \cite{Eckart:1940te,Landau1959}.
In such a framework, a relativistic fluid is characterized by the energy-momentum tensor, which admits the decomposition with respect to the fluid four-velocity\footnote{In special relativity, the fluid velocity relative to an Eulerian observer, $\vec{v}=d\vec{x}/dt$, does not constitute a complete degree of freedom. Instead, it must be replaced by a time-like normalized four-vector, defined as $u^{\mu}=dx^{\mu}/d\tau=\Lambda(v)\left(1,\vec{v}\right)$, where the fluid proper time is given by $d\tau^{2}=\eta_{\mu\nu}dx^{\mu}dx^{\nu}=dt^{2}/\Lambda(v)^{2}$ with the Minkowski metric $\eta_{\mu\nu}=\text{diag}\left(1,-1,-1,-1\right)$, and $\Lambda(v)=dt/d\tau=1/\sqrt{1-v^{2}}$ is the Lorentz factor. Furthermore, one can define a projection operator onto the space orthogonal to the four-velocity, $\Delta^{\mu}_{\nu}=\delta^{\mu}_{\nu}-u^{\mu}u_{\nu}$, thereby enabling the decomposition of a vector into parts parallel and orthogonal to the four-velocity, such as $\partial_{\mu}=u_{\mu}D+\nabla_{\mu}$ with the induced temporal derivative $D=u^{\mu}\partial_{\mu}$ and the induced spatial derivative $\nabla_{\mu}=\Delta^{\nu}_{\mu}\partial_{\nu}$.} $T^{\mu\nu}=\epsilon u^{\mu}u^{\nu}-p\Delta^{\mu\nu}+\Pi^{\mu\nu}$, where $\epsilon$ is the energy density, $p$ is the pressure, and $\Pi^{\mu\nu}$ denotes the viscous stress tensor. 
The fluid dynamic equations are governed by the conservation of the energy-momentum tensor $\partial_{\nu}T^{\mu\nu}=0$.
Specifically, projecting it onto the directions parallel and orthogonal to the fluid four-velocity yields
\begin{subequations}
	\begin{align}
		u_{\mu}\partial_{\nu}T^{\mu\nu}&=D\epsilon +\left(\epsilon +p\right)\nabla_{\mu}u^{\mu}+u_{\mu}\partial_{\nu}\Pi^{\mu\nu}=0,\\
		\Delta^{\alpha}_{\mu}\partial_{\nu}T^{\mu\nu}&=\left(\epsilon+p\right) Du^{\alpha}-\nabla^{\alpha}p+\Delta^{\alpha}_{\mu}\partial_{\nu}\Pi^{\mu\nu}=0,
	\end{align}\label{eq:1.3}
\end{subequations}
which degenerate to equations \eqref{eq:1.1a} and \eqref{eq:1.2} in the non-relativistic limit $v\leqslant 1$, respectively.
To close this Cauchy initial-value system, the constitutive relation for the viscous stress tension needs to be specified.
A phenomenological approach is built on the second law of thermodynamics, which requires that the dissipative correction results in a non-decreasing entropy production within any fluid element.
In this consideration, the zero-order correction corresponds to a relativistic ideal fluid in local equilibrium, where entropy is conserved.
The first-order correction then yields the relativistic version of the Navier–Stokes theory; however, it has been shown to be pathological, violating causality \cite{Hiscock:1987zz} and leading to instability \cite{Hiscock:1985zz}.
This issue was resolved by introducing certain second-order corrections\footnote{The M{\"u}ller-Israel-Stewart theory is commonly referred to as the second-order theory of relativistic dissipative fluid dynamics. Here, the qualifier ``second-order'' refers to the fact that the ansatz for the entropy current in this theory includes terms that are second order in the dissipative quantities. Such a requirement leads to the appearance of some second-order terms in Knudsen number in the dissipative correction. The specific derivation details are presented in subsection \ref{sec:Si}.} in the M{\"u}ller-Israel-Stewart theory \cite{Muller:1967zza,Israel:1976tn,Israel:1976efz,Israel:1979wp}, where dissipative effects appear as dynamical diffusion variables rather than instantaneous gradient terms as in the Navier–Stokes formulation.
As a result, the M{\"u}ller-Israel-Stewart theory provides a practically applicable framework for modeling relativistic dissipative fluids in high-energy environments, such as heavy-ion collisions, astrophysical phenomena, and holographic strongly coupled systems.

Fluid dynamics is an effective theory grounded in thermodynamics that captures the dynamical behavior of many-body systems near local equilibrium in the long-wavelength and low-frequency regime. 
Consequently, such a framework inevitably encounters significant challenges when dealing with interface phenomena, where microscopic effects play a crucial role.
However, interface effects play an essential role in the behavior of fluids in nature, governing a wide range of physical processes such as capillarity, phase separation, and the Rayleigh-Plateau phenomena.
This underscores the necessity of developing a theoretical framework capable of describing the dynamics of fluids with free interfaces.
A direct approach is to further modify fluid dynamics by incorporating terms that account for interface effects and to supplement it with an additional equation governing interface behavior, such as in the Young-Laplace theory \cite{Young1805,Laplace1805} and in the Cahn–Hilliard theory \cite{cahn1958free,lowengrub1998quasi}.
However, extending this strategy to a relativistic formulation poses significant difficulties.
An alternative approach is to construct a relativistic field theory with hydrodynamic properties, which permits the existence of localized steady-state configurations resembling liquid droplets, thereby providing a field theory interpretation for fluid dynamics at the phenomenological level.
Taking these considerations into account, the soliton field theory may serve as a potential candidate.

\subsection{Soliton field theory}

In field theories with certain types of nonlinearity, solitons are a class of non-dispersive solutions that can maintain their shape in a local region of space.
Coincidentally, this kind of similar matter was originally discovered in fluids.
In the first half of the nineteenth century, on the Union Canal in Scotland, the naval architect Russell observed solitary waves that maintained their shape during propagation \cite{russell1844report}, which were later interpreted as soliton solutions of the Korteweg–de Vries equation \cite{Boussinesq1877,strutt1876waves,korteweg1895xli}, used to describe waves on shallow water surfaces.
Since then, similar soliton solutions have been found to exist widely in various relativistic field theories \cite{manton2004topological}, including one-dimensional kinks, two-dimensional vortices and lumps, three-dimensional monopoles and skyrmions, as well as four-dimensional instantons, among others.
These solutions are collectively known as topological solitons, characterized by boundary conditions at spatial infinity that map onto topologically degenerate vacuum states.
Besides these, there exists a class of solutions known as non-topological solitons \cite{Lee:1991ax,Zhou:2024mea}, which approach a unique vacuum state at spatial infinity but are compelled to carry additional intrinsic conserved charge.
These particle-like solutions have demonstrated significant research value in mathematics and in multiple branches of physics, including particle physics \cite{Kusenko:1997zq}, cosmology \cite{Frieman:1988ut,Kusenko:1997si,Enqvist:1997si,Enqvist:1998en,Kusenko:2001vu,Kusenko:2008zm}, and astrophysics \cite{Lynn:1988rb}.

The present work focuses on a simple scenario of the latter, involving a single complex scalar field endowed with a global $U(1)$ symmetry and a nonlinear self-interaction, described by the following Lagrangian density:
\begin{equation}
	\mathcal{L}=\partial_{\mu}\psi\partial^{\mu}\psi^{*}-V(|\psi|),\label{eq:1.4}
\end{equation}
where the scalar potential is required to be semi-positive definite $V\geq 0$, possess a unique physical vacuum $V(0)=0$, and yield a positive mass for the matter field $m^{2}=V'(0)>0$ with $V'=dV/d|\psi|^{2}$.
The $U(1)$ symmetry of the theory naturally induces a conserved charge, typically denoted by $Q$.
Consequently, the soliton solutions therein are commonly referred to as $Q$-matter.
Such solutions are not time-independent static configurations.
Instead, they allow a separation of variables of the form $\psi(t,x)=e^{-i\omega t}\phi(x)$, where $\omega$ is the frequency and $\phi(x)$ manifests as a localized nontrivial spatial profile.
This spatial localization inherently entails the existence of an interface that separates the interior matter from the surrounding environment, which may exhibit a variety of geometries, including planar, cylindrical, spherical, or even topologically toroidal configurations \cite{Axenides:2001pi}.
Such geometric diversity not only opens a pathway to the investigation of interface phenomena but also provides a promising avenue for modeling fluid-like matter with free interfaces.

A necessary condition for the existence of such solutions is that the scalar potential satisfies $V>|\psi|^{2}V'$ within a certain range \cite{Rosen:1968mfz}.
A more stringent form of this requirement is that the potential exhibit concavity in the vicinity of the vacuum, namely $V''(0)<0$, implying a growth shallower than that of the quadratic mass term. 
Considering the following two classes of modifications to the dominant mass behavior near the vacuum:
\begin{subnumcases}{V=\label{eq:1.5}}
	m^{2}|\psi|^{2}+\lambda|\psi|^{4}+o(|\psi|^{6}), \label{eq:1.5a}\\
	m^{2}|\psi|^{2+K}+o(|\psi|^{6}),\label{eq:1.5b}
\end{subnumcases}
the existence condition requires the involved parameters to be negative, namely $\lambda,K<0$.
In this context, the high order correction in \eqref{eq:1.5a} and the exponential correction in \eqref{eq:1.5b} are respectively responsible for the existence of $Q$-matter in the gauge-mediated \cite{Kusenko:1997si} and gravity-mediated \cite{Enqvist:1997si} supersymmetry breaking scenarios in the minimal supersymmetric standard model.
Specifically, in the Affleck-Dine mechanism \cite{Affleck:1984fy} with such scalar potentials, the homogeneous scalar condensate exhibits spatial instabilities, resulting in local collapse and the formation of $Q$-matter \cite{Kasuya:1999wu,Kasuya:2000wx}.
The mechanism of this dynamical instability has long been commonly understood as arising from the negative pressure of the Affleck-Dine field.
However, there are some indications that this instability bears similarities to the sound mode instability in relativistic fluids, highlighting the hydrodynamic nature in soliton field theory \cite{Lee:1994qb,Chen:2024axd}.
Additionally, in UV-complete models, the introduction of an extra heavy or light scalar field can similarly induce a negative high order term in the polynomial potential \cite{Heeck:2022iky}, thereby ensuring the emergence of $Q$-matter.

From a conceptual standpoint, fluids and fields are fundamentally distinct physical systems: fluids represent a averaged description of microscopic particles, whereas fields are more fundamental constructs.
Establishing a duality between the two is an elegant yet challenging endeavor \cite{Crossley:2015evo}.
Currently, the correspondence between a real scalar field and a relativistic fluid has been widely recognized \cite{Madsen:1988ph} and extensively applied in cosmological researches \cite{Weinberg:2008zzc}.
In the case of minimal coupling to gravity, a real scalar field with a time-like gradient identified as the four-velocity is equivalent to an ideal fluid, whereas dissipative effects arise under non-minimal coupling to the Ricci curvature.
However, this duality mainly manifests as a formal similarity in the structure of the energy-momentum tensor, rather than in the dispersion relation.
Essentially, a real scalar field does not possess hydrodynamic modes, let alone the characteristic sound mode of fluids.
The duality in dispersion relation instead emerges within the framework of complex scalar field theory \cite{Suarez:2015fga}.
In view of the presence of sound modes in complex scalar fields, one of the results of this work is to demonstrate that the mechanism of soliton formation is a phase separation process induced by the sound mode instability.
Furthermore, this study extends the findings of the work \cite{Chen:2024axd} by providing further evidence that the interface of a soliton resembles a fluid membrane subject to surface tension.
In this way, the soliton field theory provides a unified platform for modeling not only the sound mode behavior in relativistic fluids but also the interface phenomena driven by surface tension.
Similar efforts have been made in the superfluid bubble model \cite{Armas:2015ssd,Armas:2016xxg}, the holographic first order phase transition model \cite{Attems:2018gou,Bea:2022mfb}, and related approaches.

The organization of the paper is as follows.
In section \ref{sec:Hi}, we introduce two types of long-wavelength dynamical instabilities in fluids.
First, we derive the dispersion relation for a relativistic dissipative fluid, which reveals the sound mode instability induced by thermodynamic effects.
Second, we derive the dispersion relation for a liquid column in the Rayleigh-Plateau model, uncovering the membrane instability driven by surface tension.
Thereafter, we demonstrate that both types of hydrodynamic instabilities also arise in the soliton field theory.
First, in section \ref{sec:Sis}, we show that the dispersion relation of the soliton field contains the sound mode that exhibits dynamical instability in regimes of thermodynamic instability, thereby leading to the formation of solitons or vacuum bubbles through phase separation.
Second, in section \ref{sec:Mi}, we reveal that the interface pressure of the soliton satisfies a Young-Laplace-type relation, causing soliton configurations with cylindrical interfaces to undergo the membrane instability analogous to the Rayleigh-Plateau phenomenon.
Finally, we conclude in section \ref{sec:C} with a summary and outlook.


\section{Hydrodynamic instabilities}\label{sec:Hi}
In this section, we introduce two types of dynamical instabilities dominated by hydrodynamic modes in the long-wavelength regime.
One type is determined by the intrinsic thermodynamic properties of fluids and is associated with phase separation driven by a first-order thermodynamic phase transition.
The other type arises from interface effects and serves as the primary mechanism behind the topological transition in the Rayleigh-Plateau phenomenon.

\subsection{Sound mode instability}\label{sec:Si}
To match the soliton field theory with a conserved charge, we consider a relativistic fluid endowed with an additional conserved particle number, described by the following particle current and energy-momentum tensor:
\begin{subequations}
	\begin{align}
		N^{\mu}&=nu^{\mu}+n^{\mu},\\
		T^{\mu\nu}&=\epsilon u^{\mu}u^{\nu}-p\Delta^{\mu\nu}+\Pi^{\mu\nu}+q^{\mu}u^{\nu}+q^{\nu}u^{\mu},
	\end{align}
\end{subequations}
where $n$ denotes the particle number density.
The dissipative quantities involved, namely the particle diffusion current $n^{\mu}$, the energy diffusion current $q^{\mu}$, and the viscous stress tensor $\Pi^{\mu\nu}$, are all orthogonal to the fluid four-velocity
\begin{equation}
	n^{\mu}u_{\mu}=0,\quad q^{\mu}u_{\mu}=0,\quad \Pi^{\mu\nu}u_{\mu}=0.\label{eq:2.2}
\end{equation}
For convenience, the viscous stress tensor is typically decomposed into a traceless part and a trace part
\begin{equation}
	\Pi^{\mu\nu}=\pi^{\mu\nu}-\Pi\Delta^{\mu\nu},
\end{equation}
with the shear stress tensor $\pi^{\mu\nu}$ and the bulk viscous pressure $\Pi$.
Additionally, several key fluid kinematic variables are introduced: 

\noindent the four-acceleration
\begin{equation}
	a_{\mu}=Du_{\mu},
\end{equation}

\noindent the vorticity tensor
\begin{equation}
	\chi_{\mu\nu}=\Delta^{\alpha}_{\mu}\Delta^{\beta}_{\nu}\left(\partial_{\alpha}u_{\beta}-\partial_{\beta}u_{\alpha}\right)/2,
\end{equation}

\noindent and the expansion tensor
\begin{equation}
	\theta_{\mu\nu}=\Delta^{\alpha}_{\mu}\Delta^{\beta}_{\nu}\left(\partial_{\alpha}u_{\beta}+\partial_{\beta}u_{\alpha}\right)/2=\sigma_{\mu\nu}+\frac{1}{3}\theta\Delta_{\mu\nu},
\end{equation}
with the shear tensor $\sigma_{\mu\nu}$ (traceless part) and the volume expansion $\theta$ (trace part).
With these definitions, the fluid dynamic equations derived from the conservation of the particle current $\partial_{\mu}N^{\mu}=0$ and the conservation of the energy-momentum tensor $\partial_{\nu}T^{\mu\nu}=0$ take the following form:
\begin{subequations}
	\begin{align}
		0&=Dn+\theta n+\left(\nabla-a\right)_{\mu}n^{\mu},\\
		0&=D\epsilon +\left(\epsilon +p+\Pi\right)\theta-\pi^{\mu\nu}\sigma_{\mu\nu}+\left(\nabla-2a\right)_{\mu}q^{\mu},\\
		0&=\left(\epsilon+p+\Pi\right) a^{\lambda}-\nabla^{\lambda}\left(p+\Pi\right)+\Delta^{\lambda}_{\mu}\left(\nabla-a\right)_{\nu}\pi^{\mu\nu}+\left[\Delta\left(D+\frac{4}{3}\theta\right)+\chi+\sigma\right]^{\mu\lambda}q_{\mu}.
	\end{align}\label{eq:2.7}
\end{subequations}
The above set of equations can be regarded as the evolution equations for the primary fluid dynamical quantities $\{n,\epsilon,u^{\mu}\}$.
To close the system, it is necessary to further specify constitutive relations for the dissipative quantities $\{\Pi,n^{\mu},q^{\mu},\pi^{\mu\nu}\}$.
One starting point lies in the freedom of defining the fluid four-velocity, which is typically endowed with specific physical meaning through one of two common choices.
The first identifies it with the particle current $N^{\mu}=nu^{\mu}$, leading to $n^{\mu}=0$, a formulation known as the Eckart frame \cite{Eckart:1940te}.
The other aligns it with the energy current $u_{\nu}T^{\mu\nu}=\epsilon u^{\mu}$, resulting in $q^{\mu}=0$, commonly referred to as the Landau-Lifshitz frame \cite{Landau1959}.
The physics is independent of the choice of frame.
In this work, we adopt the Eckart frame, as it provides an explicit definition of the fluid four-velocity $u^{\mu}=N^{\mu}/\sqrt{N^{\alpha}N_{\alpha}}$.
The fluid dynamic formulation in the Landau-Lifshitz frame is presented in the Appendix \ref{sec:Aa}.

Given the vanishing particle diffusion current $n^{\mu}=0$, the constitutive relations for the remaining dissipative quantities $\{\Pi,q^{\mu},\pi^{\mu\nu}\}$ are typically constructed on phenomenological grounds to ensure compliance with the second law of thermodynamics, that is, dissipative corrections result in an increase in entropy within any fluid element.
To this end, the entropy four-current is introduced
\begin{equation}
	S^{\mu}=su^{\mu}+F^{\mu},
\end{equation}
where $s$ is the entropy density and $F^{\mu}$ denotes the dissipation-induced correction part.
Combining the basic equilibrium thermodynamic relations
\begin{equation}
	\epsilon+p=Ts+\mu n,\quad d\epsilon=Tds+\mu dn,
\end{equation}
with the temperature $T$ and the chemical potential $\mu$, one obtains the following expression for the entropy production rate
\begin{equation}
	\partial_{\mu}S^{\mu}=\partial_{\mu}F^{\mu}-T^{-1}\left[\Pi\theta+\left(\nabla-2a\right)_{\mu}q^{\mu}-\pi^{\mu\nu}\sigma_{\mu\nu}\right].
\end{equation}
Once the form of the correction term $F^{\mu}$ is specified, the relations between the dissipative quantities and the gradients of the primary fluid dynamical quantities are constrained by the requirement of entropy production $\partial_{\mu}S^{\mu}\geq 0$.
Since the dissipative quantities are treated as small corrections, a natural approach is to expand the non-equilibrium entropy current in powers of these quantities.
In the relativistic framework, such an expansion is retained to at least second order \cite{Muller:1967zza,Israel:1976tn,Israel:1976efz,Israel:1979wp}:
\begin{equation}
	TF^{\gamma}=\left(\alpha_{1}\Pi+\alpha_{\Pi}\Pi^{2}-\alpha_{q}q_{\mu}q^{\mu}+\alpha_{\pi}\pi_{\mu\nu}\pi^{\mu\nu}\right)u^{\gamma}+\beta_{1} q^{\gamma}+\beta_{\Pi}\Pi q^{\gamma}+\beta_{\pi}\pi^{\mu\gamma}q_{\mu},\label{eq:2.11}
\end{equation}
with expansion coefficients $\{\alpha_{i},\beta_{i}\}$, giving rise to the following form of the entropy production rate
\begin{equation}
	\partial_{\mu}S^{\mu}=T^{-1}\alpha_{1}D\Pi+T^{-1}\left(\beta_{1}-1\right) \left(\nabla-a\right)_{\mu}q^{\mu}+\Pi f_{\Pi}-q_{\mu}f^{\mu}_{q}+\pi_{\mu\nu}f^{\mu\nu}_{\pi}.
\end{equation}
To ensure compliance with the second law of thermodynamics, a minimal requirement is that the above expression for entropy production be a semi-positive definite function of the dissipative quantities, leading to the conditions $\alpha_{1}=0$ and $\beta_{1}=1$, with a linear relation between the extended thermodynamic forces and the dissipative fluxes
\begin{equation}
	f_{\Pi}=\frac{\Pi}{c_{\Pi} T},\quad f^{\mu}_{q}=\frac{q^{\mu}}{c_{q} T},\quad f^{\mu\nu}_{\pi}=\frac{\pi^{\mu\nu}}{c_{\pi}T},\label{eq:2.13}
\end{equation}
with positive proportionality coefficients: bulk viscosity $c_{\Pi}$, thermal conductivity $c_{q}$, and shear viscosity $c_{\pi}$.
Defining the relaxation times
\begin{equation}
	\tau_{\Pi}=-2c_{\Pi} \alpha_{\Pi},\quad \tau_{q}=-2c_{q} \alpha_{q},\quad \tau_{\pi}=-2c_{\pi} \alpha_{\pi},\label{eq:2.14}
\end{equation}
the relations \eqref{eq:2.13} yield the following relaxation-type constitutive equations for the dissipative quantities:
\begin{subequations}
	\begin{align}
		\tau_{\Pi}D\Pi+\Pi&=-c_{\Pi}\theta+\delta_{\Pi},\\
		\tau_{q}\Delta^{\mu}_{\nu}Dq^{\nu}+q^{\mu}&=c_{q}\left(\nabla\ln T-a\right)^{\mu} +\delta^{\mu}_{q},\\
		\tau_{\pi}\Delta^{\mu}_{\alpha}\Delta^{\nu}_{\beta} D\pi^{\alpha\beta}+\pi^{\mu\nu}&=c_{\pi}\sigma^{\mu\nu}+\delta^{\mu\nu}_{\pi},
	\end{align}\label{eq:2.15}
\end{subequations}
where the first terms on the right-hand side represent the first-order correction in the Knudsen number, corresponding to the Navier-Stokes contribution, and the second terms consist of second-order corrections in the Knudsen number:
\begin{subequations}
	\begin{align}
		\delta_{\Pi}&=-\frac{\tau_{\Pi}}{2}\left[\theta+D\ln\left(\frac{\tau_{\Pi}}{c_{\Pi} T}\right)\right]\Pi+c_{\Pi} \beta_{\Pi}\left[\nabla- c_{1}a+c_{2}\nabla\ln\left(\beta_{\Pi}/T\right)\right]_{\mu}q^{\mu},\\
		\delta^{\mu}_{q}&=-\frac{\tau_{q}}{2}\left[\theta+D\ln\left(\frac{\tau_{q}}{c_{q} T}\right)\right]q^{\mu}-c_{q} \beta_{\Pi}\left[\nabla- \left(1-c_{1}\right)a+\left(1-c_{2}\right)\nabla\ln\left(\beta_{\Pi}/T\right)\right]^{\mu}\Pi\nonumber\\
		&\quad-c_{q} \beta_{\pi}\Delta^{\mu}_{\alpha}\left[\nabla-\left(1-c_{3}\right)a+\left(1-c_{4}\right)\nabla\ln\left(\beta_{\pi}/T\right)\right]_{\nu}\pi^{\alpha\nu},\\
		\delta^{\mu\nu}_{\pi}&=-\frac{\tau_{\pi}}{2}\left[\theta+D\ln\left(\frac{\tau_{\pi}}{c_{\pi} T}\right)\right]\pi^{\mu\nu}+c_{\pi} \beta_{\pi}\left[\Delta^{\mu}_{\alpha}\Delta^{\nu}_{\beta}\left[\nabla-c_{3}a +c_{4}\nabla\ln\left(\beta_{\pi}/T\right)\right]^{(\alpha}q^{\beta)}\right.\nonumber\\
		&\quad\left.-\left[\nabla-c_{3}a+c_{4}\nabla\ln\left(\beta_{\pi}/T\right)\right]_{\alpha}q^{\alpha}\Delta^{\mu\nu}/3\right],
	\end{align}
\end{subequations}
with arbitrary constants $\{c_{i}\}$, which arise from the ambiguity in assigning cross terms.
Due to the requirement of hyperbolicity, the relaxation times \eqref{eq:2.14} must be positive, implying that the entropy density in a non-equilibrium state is lower than that in equilibrium
\begin{equation}
	u_{\mu}S^{\mu}=s-\frac{1}{2T}\left(\frac{\tau_{\Pi}}{c_{\Pi}}\Pi^{2}-\frac{\tau_{q}}{c_{q}}q_{\mu}q^{\mu}+\frac{\tau_{\pi}}{c_{\pi}}\pi_{\mu\nu}\pi^{\mu\nu}\right)\leq s,\label{eq:2.17}
\end{equation}
consistent with the thermodynamic principle that equilibrium states maximize entropy.

In the M{\"u}ller-Israel-Stewart theory, the second-order corrections to the entropy four-current endows the dissipative quantities with diffusive dynamical behavior, which reduces to the Navier-Stokes theory at first order.
In that case, the dissipative quantities appear as gradients of the primary fluid dynamical quantities, a feature that, as revealed by linear perturbation theory, leads to a violation of causality and the emergence of instability.
Let us consider the perturbations along the spatial $x$-direction around a state in equilibrium and at rest:
\begin{equation}
	\begin{aligned}
		n&=n_{0}+\delta_{x} n,\quad& \epsilon&=\epsilon_{0}+\delta_{x} \epsilon, \quad& u^{\mu}&=\left(1,\vec{0}\right)+\left(0,\delta_{x} u^{i}\right),\\
		\Pi&=\delta_{x} \Pi,\quad& q^{\mu}&=\left(0,\delta_{x}q^{i}\right), \quad& \pi^{\mu\nu}&=\left[\left(0,\vec{0}\right),\left(\vec{0},\delta_{x}\pi^{ij}\right)\right],
	\end{aligned}\label{eq:2.18}
\end{equation}
with the spatial perturbation operator $\delta_{x}=e^{-i\Omega t-ikx}\delta$.
The real part of the perturbation frequency Re$[\Omega]$ determines the propagation speed of the mode, while the imaginary part Im$[\Omega]$ governs its decay (Im$[\Omega]<0$) or growth (Im$[\Omega]>0$) rate.
The wavelength associated with the perturbation is given by $\lambda=2\pi/k$.
Due to the normalization of the four-velocity $u^{\mu}u_{\mu}=1$ and the orthogonality of the dissipative quantities \eqref{eq:2.2}, the perturbations in their temporal components are of higher-order smallness.
Substituting the perturbation ansatz \eqref{eq:2.18} into the fluid dynamic equations \eqref{eq:2.7} and the constitutive equations \eqref{eq:2.15} and truncating the expansion at first order, with the equations of state $p=p(n,\epsilon)$ and $T=T(n,\epsilon)$, one obtains the following perturbation equations with a simple algebraic form:
\begin{equation}
	\text{diag}\left\lbrace M_{l},M_{t},M_{t},M_{d}\right\rbrace\left(\delta A^{x},\delta A^{y},\delta A^{z},\delta A^{d}\right)^{T}=0,\label{eq:2.19}
\end{equation}
with $\delta A^{x}=\{\delta n,\delta\epsilon,\delta\Pi,\delta u^{x},\delta q^{x},\delta\pi^{xx}\}$, $\delta A^{(y,z)}=\{\delta u^{(y,z)},\delta q^{(y,z)},\delta\pi^{x(y,z)}\}$ and $\delta A^{d}=\{\delta\pi^{yz},\delta\pi^{yy}-\delta\pi^{zz}\}$.
The existence of non-trivial solutions requires the determinant of the coefficient matrix to vanish, namely
\begin{equation}
	\det\left(M_{l}\right)\det\left(M_{t}\right)^{2}\det\left(M_{d}\right)=0,
\end{equation}
which respectively yields the following three types of dispersion relations:

\noindent the longitudinal dispersion relation
\begin{equation}
	\begin{aligned}
		&0=\left[\Omega - \left(h^{-1}_{0}\frac{\partial p}{\partial n}+\frac{\partial p}{\partial\epsilon}\right)\frac{k^{2}}{\Omega}-\frac{K_{\Pi}+K_{\pi}}{\epsilon_{0}+p_{0}}\right]\left[c^{-1}_{q}\left(\tau_{q}\Omega+i\right) -\frac{\partial\ln T}{\partial\epsilon}\frac{k^{2}}{\Omega}-\beta^{2}_{\Pi}K_{\Pi}-\beta^{2}_{\pi}K_{\pi}\right]\\
		&-\left(\Omega+\beta_{\Pi}K_{\Pi}-\beta_{\pi}K_{\pi}-\frac{\partial p}{\partial\epsilon}\frac{k^{2}}{\Omega}\right)\left[\frac{\Omega+\beta_{\Pi}K_{\Pi}-\beta_{\pi}K_{\pi}}{\epsilon_{0} +p_{0}}-\left( h^{-1}_{0}\frac{\partial\ln T}{\partial n}+\frac{\partial\ln T}{\partial\epsilon}\right)\frac{k^{2}}{\Omega}\right],
	\end{aligned}\label{eq:2.21}
\end{equation}

\noindent the transverse dispersion relation
\begin{equation}
	0=\left[\left(\epsilon_{0}+p_{0}\right)\Omega-\frac{3}{4}K_{\pi}\right]\left[c^{-1}_{q}\left(\tau_{q}\Omega+i\right) -\frac{3}{4}\beta^{2}_{\pi}K_{\pi}\right]-\left(\Omega-\frac{3}{4}\beta_{\pi}K_{\pi}\right)^{2},\label{eq:2.22}
\end{equation}

\noindent and the damped dispersion relation
\begin{equation}
	0=\tau_{\pi}\Omega+i,\label{eq:2.23}
\end{equation}
where $h_{0}=\frac{\epsilon_{0} +p_{0}}{n_{0}}$ is the enthalpy per particle, and the auxiliary quantities $\{K_{\Pi},K_{\pi}\}$ are defined as $K_{\Pi}=\frac{c_{\Pi}k^{2}}{\tau_{\Pi}\Omega+i},K_{\pi}=\frac{2}{3}\frac{c_{\pi}k^{2}}{\tau_{\pi}\Omega+i}$.
The dispersion relation \eqref{eq:2.23} indicates that the associated mode decays exponentially with a constant damping rate $\tau^{-1}_{\pi}$, independent of the wavelength.
Similar strongly damped modes arise in the other two dispersion branches as well.
In this work, we do not pursue such modes further, as they are expected to have negligible impact on physical observations.
Instead, our primary interest lies in the hydrodynamic modes in the long-wavelength regime, which are defined by the condition $\Omega(k\rightarrow 0)$=0. 

The inherent instability of the relativistic Navier-Stokes theory is encoded in the transverse dispersion relation \eqref{eq:2.22}, which gives rise to a hydrodynamic mode and a non-hydrodynamic mode in the long-wavelength regime
\begin{subequations}
	\begin{align}
		\Omega_{\pi}&=-\frac{1}{2}\frac{c_{\pi}i}{\epsilon_{0}+p_{0}}k^{2}+o(k^{4}),\\
		\Omega_{d}&=-\frac{\left(\epsilon_{0}+p_{0}\right)i}{\tau_{q}\left(\epsilon_{0}+p_{0}\right)-c_{q}}+o(k^{2}).
	\end{align}\label{eq:2.24}
\end{subequations}
The hydrodynamic mode $\Omega_{\pi}$ exhibits damped behavior, characterizing the viscous attenuation of transverse shear stresses.
The issue arises with the non-hydrodynamic mode $\Omega_{d}$, which displays strong damping in the M{\"u}ller-Israel-Stewart theory with the condition $\tau_{q}\left(\epsilon_{0}+p_{0}\right)-c_{q}>0$, but becomes dynamically unstable in the Navier-Stokes limit $\tau_{q}=0$, manifesting as exponential growth.
At that point, the transverse sector of the perturbation equations can be reduced to an elliptic equation
\begin{equation}
	M^{t}\delta A^{(y,z)}=c_{q}\partial^{2}_{t}\delta_{x}u^{(y,z)}-\left(\epsilon_{0}+p_{0}\right)\partial_{t}\delta_{x}u^{(y,z)}+\frac{1}{2}c_{\pi}\partial^{2}_{x}\delta_{x} u^{(y,z)}=0,
\end{equation}
thereby violating causality.
This is precisely where the pathological nature of the relativistic Navier-Stokes theory lies.

In self-consistent frameworks, such as the M{\"u}ller-Israel-Stewart theory or the ideal fluid limit, there exists a dynamical instability governed by hydrodynamic modes.
This instability does not stem from any pathology of the theory itself, but rather originates from the intrinsic properties of the fluid, representing a genuine and observable physical phenomenon.
The triggering condition is encoded in the longitudinal dispersion relation \eqref{eq:2.21}, which yields the following two hydrodynamic modes in the long-wavelength regime:
\begin{subequations}
	\begin{align}
		\Omega_{q}&= - \frac{c_{q}i}{nTc_{\hat{s}}}k^{2}+o(k^{4}),\\
		\Omega_{s}&=v_{s}k-i\Gamma k^{2}+o(k^{3}),
	\end{align}
\end{subequations}
with the isobaric specific heats
\begin{equation}
	c_{\hat{s}}=T\left[\frac{\partial\hat{s}}{\partial T}\right]_{p},\quad c_{\epsilon}=\left[\frac{\partial\epsilon}{\partial T}\right]_{p},
\end{equation}
the adiabatic speed of sound, and the sound attenuation coefficient
\begin{equation}
	v^{2}_{s} =\left[\frac{\partial p}{\partial \epsilon}\right]_{\hat{s}},\quad \Gamma=\frac{1}{2\left(\epsilon_{0}+p_{0}\right)}\left[c_{\Pi}+\frac{2}{3}c_{\pi}+\left(\frac{v_{s}c_{\epsilon}}{nc_{\hat{s}}}\right)^{2}c_{q}\right],
\end{equation}
where $\hat{s}=s/n$ is the entropy density per particle.
The derivation involves some thermodynamic relations, which are presented in the Appendix \ref{sec:Tr}.
The first hydrodynamic mode $\Omega_{q}$ exhibits damped behavior under the condition of positive specific heat, $c_{\hat{s}}>0$, characterizing the thermal diffusion of temperature fluctuations.
The second hydrodynamic mode $\Omega_{s}$, commonly referred to as the sound mode, remains dynamically stable provided that the squared speed of sound is positive $v^{2}_{s}>0$.
Such a mode propagates with phase and group velocities equal to $v_{s}$, and is accompanied by damped behavior contributed by the positive-definite sound attenuation coefficient $\Gamma$ at the subleading order.
A violation of either of these two positivity conditions can render the corresponding hydrodynamic mode dynamically unstable, leading to the emergence of perturbations with negative energy \cite{Hiscock:1983zz}.
We focus solely on the instability of the sound mode, as it can arise even in the ideal fluid limit, unlike the other one.
In this case, the speed of sound becomes imaginary, $v_{s}=\pm\sqrt{|\left[\partial p/\partial\epsilon\right]_{\hat{s}}|}i$, with the positive branch indicating a dynamical instability characterized by an exponential growth rate Im$[\Omega_{s}]=|v_{s}|k+o(k^{2})$.
This reveals an intriguing phenomenon: a thermodynamic instability, $\left[\partial p/\partial\epsilon\right]_{\hat{s}}<0$, corresponds to a dynamical instability.
Although the sound attenuation coefficient $\Gamma$ becomes formally indefinite under such condition, it remains positive in a system where viscous effects dominate over thermal conduction.
As a result, short-wavelength perturbations are effectively damped by viscous effects, introducing a characteristic threshold $k_{c}=|v_{s}|/\Gamma$ that confines the instability to the long-wavelength regime $0<k<k_{c}$, or $\lambda>2\pi/k_{c}$.
In the ideal fluid limit, such a threshold disappears, rendering the instability active across the entire wavelength spectrum and more pronounced at short wavelengths due to an enhanced growth rate.

\begin{figure}
	\begin{center}
		\subfigure[]{\includegraphics[width=.492\linewidth]{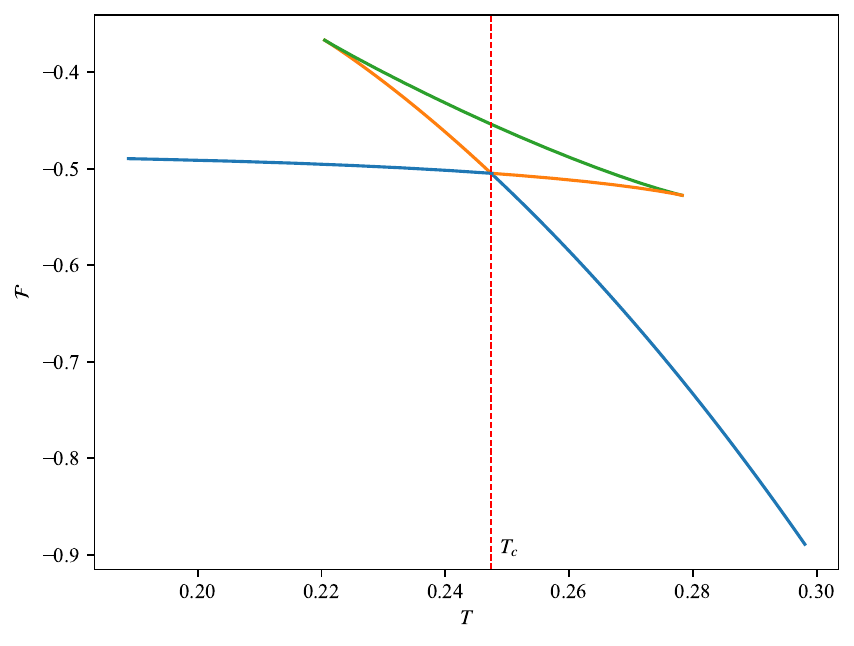}\label{fig:3}}
		\subfigure[]{\includegraphics[width=.48\linewidth]{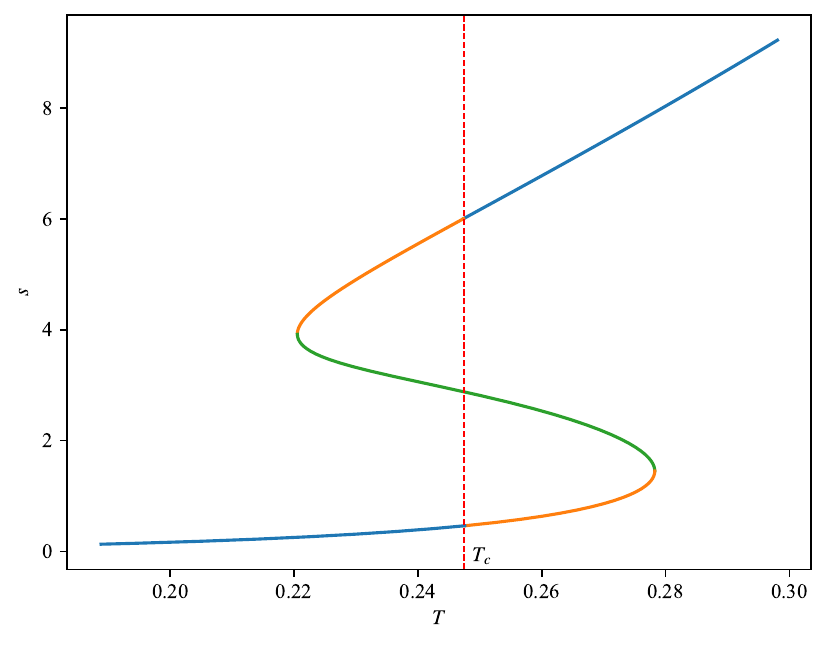}\label{fig:4}}
		\caption{Phase diagrams of holographic fluids with a thermal first-order phase transition: free energy density (a) and entropy density (b) as functions of temperature.}
		\label{fig:3-4}
	\end{center}
\end{figure}

This type of hydrodynamic instability can break spatial translational invariance and is expected to dynamically trigger a thermodynamic phase transition, wherein the thermodynamic properties of the fluid undergo drastic changes, culminating in the emergence of new thermal phases with thermodynamic stability $\left[\partial p/\partial\epsilon\right]_{\hat{s}}>0$.
To provide an intuitive picture of this instability, we take an example of holographic fluids without particle current $N^{\mu}=0$, which have been employed to investigate the physical properties of strongly coupled QCD-like systems within the framework of AdS/CFT theory \cite{Gubser:2008ny}.
In this case, the thermodynamic instability is equivalent to a negative specific heat\footnote{For systems without a particle current, the fluid four-velocity is not well-defined in the Eckart formalism, and one may instead work in the Landau-Lifshitz frame, as done in the appendix \ref{sec:Aa}. The absence of particle flow eliminates thermal diffusion, and thus no hydrodynamic mode analogous to $\Omega_{q}$ arises. Consequently, the negative specific heat does not trigger an instability associated with thermal diffusion, but instead gives rise to an imaginary speed of sound, inducing an instability in the sound mode.}
\begin{equation}
	\frac{d p}{d \epsilon}=\frac{sdT}{Tds}=\frac{s}{c_{s}}<0.
\end{equation}
Figure \ref{fig:3-4} illustrates the thermodynamic relations in the presence of a thermal first-order phase transition, marked by a discontinuity in the first derivative of the branch with the lowest free energy at the transition temperature $T_{c}$.
The regions marked by different colors represent thermal phases with distinct properties.
Among them, the blue regions correspond to the ground states, which can be intuitively understood as analogous to ice and water.
The orange regions represent the superheated and supercooled states, respectively.
Intriguingly, the green region, known as the spinodal region, is characterized by a negative specific heat.
As a result, the sound mode instability occurs \cite{Janik:2015iry}.
In real-time dynamics \cite{Janik:2017ykj}, the spatial translational symmetry is broken, the local thermodynamic properties undergo transitions, and ultimately phase separation occurs accompanied by the formation of interfaces.
The final state is a two-component fluid, manifested as different spatial regions occupied by two ground states at the transition temperature with the same free energy but different entropies, connected by a domain wall.
This is an ice-water mixture!

In section \ref{sec:Sis}, we demonstrate that the homogeneous soliton field also exhibits the sound mode instability induced by thermodynamic instability in the long-wavelength regime of perturbations.
This instability drives the local collapse of the system, leading to the formation of particle-like solitons as a new thermal phase with thermodynamic stability, which possesses a interface isolating it from the external environment.
This physical phenomenon can be regarded as a phase separation process, with the final state analogous to the above two-component fluid.

\subsection{Membrane instability}
The sound mode instability is governed by the intrinsic thermodynamic properties of the fluid and is independent of the characteristics of the interface.
This type of instability, when dynamically driving a thermal phase transition, is typically accompanied by the formation of interfaces, whose geometric configuration depends on the symmetry of the applied perturbation, such as planar, cylindrical, spherical, or other topological structure.
In nature, planar and spherical fluid membranes are commonly observed, such as lake surfaces and raindrops, whereas fluid membranes that are locally cylindrical are rarely encountered.
A similar phenomenon arises in the world of solitons.
From the perspective of the dynamics induced by sound mode instability, cylindrical domain walls are not in any way special and can be dynamically generated by imposing perturbations with the corresponding cylindrical symmetry.
This absence, therefore, may signal a deeper underlying principle.

In hindsight, one may interpret that cylindrical fluid membranes are dynamically unstable and tend to evolve into other geometric structures under perturbations.
This intuition can be supported by a familiar kitchen example: the end of a cylindrical stream of water flowing from a faucet inevitably breaks into droplets.
Furthermore, we may draw the empirical observation that the thinner and longer the stream, the more susceptible it is to fragmentation.
This is the well-known Rayleigh-Plateau phenomenon.
There has been a long history of relevant research.
The earliest scientific record may be traced back to Mariotte's description of the splitting of a stream of water from a container in the end of the seventeenth century, a phenomenon attributed to some kind of external force \cite{mariotte1700}.
A more scientifically grounded investigation began in the early nineteenth century, when Savart systematically investigated the factors influencing the breakup of fluid jets in experiments, showing signs that the cause might be internal \cite{Savart:1833}.
At that time, this internal cause, surface tension, had actually already been mathematically introduced by Young and Laplace a few decades earlier to study the curvature of fluid surfaces \cite{Young1805,Laplace1805}, which was later developed into the well-known Young-Laplace equation
\begin{equation}
	p=\gamma\left(R^{-1}_{1}+R^{-1}_{2}\right),\label{eq:2.30}
\end{equation}
where $p$ denotes the pressure across the fluid interface induced by curvature, $\gamma$ is the surface tension, and $R_{1},R_{2}$ are the principal radii of curvature.
Eventually, in the mid-nineteenth century, Plateau considered the role of surface tension in jet breakup \cite{Plateau:1849}, and the insights were later confirmed by Rayleigh through theoretical modeling \cite{Rayleigh1878}.
To this day, the Rayleigh-Plateau phenomenon has found wide-ranging applications and continues to be a subject of active research in modern disciplines, such as physics, chemistry, biology, and materials science.

\begin{figure}
	\centering\includegraphics[width=.55\linewidth]{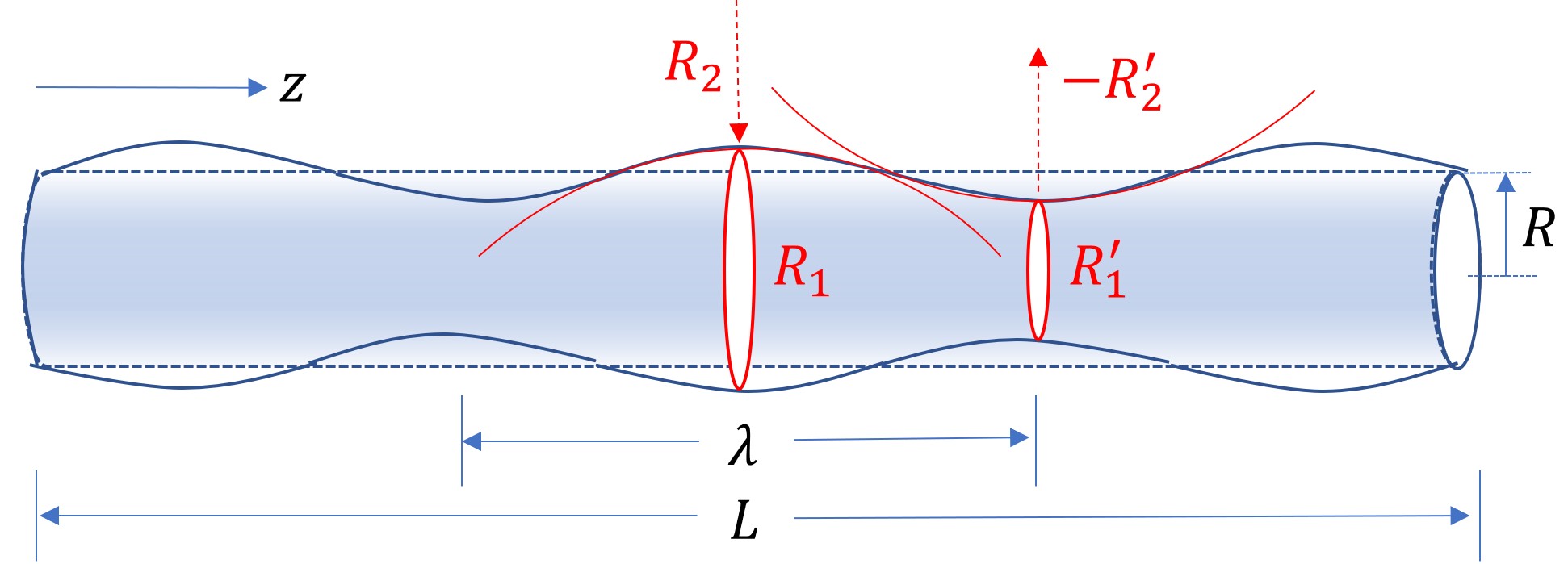}
	\caption{Schematic diagram of the Rayleigh-Plateau model with a water column of length $L$ and radius $R$ being affected by a perturbation of wavelength $\lambda$. $R_{1}$ and $R_{2}$ denote the principal radii of curvature in the cross-sectional and axial directions of the cylinder, respectively.}
	\label{fig:5}
\end{figure}

We now turn to a theoretical derivation of the Rayleigh-Plateau instability using the classical ideal fluid dynamic equations in conjunction with the Young-Laplace equation.
For simplicity, we neglect the effects of viscosity, which were absent in the original analysis but subsequently addressed in Rayleigh's follow-up study \cite{Rayleigh1892}.
Figure \ref{fig:5} provides an intuitive schematic of the Rayleigh-Plateau model.
Since the radius of curvature in the cross-sectional direction, which produces an inward pressure, at the trough is smaller than that at the crest, $R'_{1}<R_{1}$, the Young-Laplace equation \eqref{eq:2.30} indicates that this results in a pressure that drives the fluid from the trough toward the crest, causing the neck to shrink further and indicating instability.
On the other hand, the radius of curvature along the axial direction is positive at the crest and negative at the trough, contributing an opposing effect.
Therefore, the Rayleigh-Plateau instability can essentially be viewed as a competition between the two principal radii of curvature.
The triggering condition is that the pressure contribution from the curvature radius in the cross-sectional direction dominates, characterized by $R\ll\lambda$, which explains why slender fluid jets are more prone to instability.
Quantitatively, by imposing a perturbation
\begin{equation}
	R=R_{0}+e^{-i\Omega t-ikz-il\varphi}\delta R,
\end{equation}
on a cylinder with equilibrium radius $R_{0}$, where $l$ is an integer characterizing the angular excitation, the principal radii of curvature are obtained as
\begin{subequations}
	\begin{align}
		R^{-1}_{1}&=R^{-1}_{0}+\left(l^{2}-1\right) R^{-2}_{0}e^{-i\Omega t-ikz-il\varphi}\delta R+o(\delta^{2}),\\
		R^{-1}_{2}&=k^{2}e^{-i\Omega t-ikz-il\varphi}\delta R+o(\delta^{2}).
	\end{align}
\end{subequations}
From the Young-Laplace equation \eqref{eq:2.30}, the perturbation induces a pressure variation at the cylindrical surface given by
\begin{equation}
	p\left(R_{0}\right)-p_{0}=\gamma R^{-2}_{0}\left(l^{2}+k^{2}R^{2}_{0}-1\right)\delta Re^{-i\Omega t-ikz-il\varphi},\label{eq:2.33}
\end{equation}
with the equilibrium pressure $p_{0}=\gamma R^{-1}_{0}$.
The occurrence of instability requires that the pressure at the trough be inward $\delta p>0$, and at the crest be outward $\delta p<0$, which implies that the term inside the parentheses must be negative
\begin{equation}
	l^{2}+k^{2}R^{2}_{0}<1.
\end{equation}
Therefore, only perturbations with axisymmetry $l=0$ and long wavelength $\lambda>\lambda_{c}=2\pi R_{0}$ can trigger the Rayleigh-Plateau instability. 
It can now be concluded that a liquid column with length greater than its cross-sectional circumference is dynamically unstable, undergoing breakup under axisymmetric perturbations with wavelength $\lambda_{c}<\lambda<L$.

The complete dispersion relation can be obtained by solving the classical ideal fluid dynamic equations with appropriate boundary conditions.
To this end, an axisymmetric perturbation is imposed around a state in equilibrium and at rest
\begin{equation}
	\vec{v}=e^{-i\Omega t-ikz}\left(\delta v^{r},\delta v^{\varphi},\delta v^{z}\right),\quad p=p_{0}+e^{-i\Omega t-ikz}\delta p,
\end{equation}
which, when substituted into the continuity and Euler equations \eqref{eq:1.1} and assuming a constant fluid density, yields the following master equation:
\begin{equation}
	r^{2}\frac{d^{2}}{dr^{2}}\delta p+r\frac{d}{dr}\delta p-k^{2}r^{2}\delta p=0,
\end{equation}
an modified Bessel equation.
The general solution is
\begin{equation}
	\delta p=C_{1}I_{0}\left(kr\right)+C_{2}K_{0}\left(kr\right),
\end{equation}
where $I_{0}$ and $K_{0}$ are the modified Bessel functions of the first and second types, respectively.
Since the branch of the second type diverges at the origin $K_{0}\left(0\right)\rightarrow\infty$, we must set $C_{2}=0$.
The other coefficient can be determined from the boundary condition \eqref{eq:2.33}, yielding
\begin{equation}
	C_{1}=\gamma R^{-2}_{0}\left(k^{2}R^{2}_{0}-1\right)\delta RI^{-1}_{0}\left(kR_{0}\right).
\end{equation}
From the Euler equation, one can further obtain the perturbation of the radial component of the fluid velocity
\begin{equation}
	\delta v^{r}=-\frac{i}{\rho\Omega}\frac{d}{dr}\delta p=-\frac{ikC_{1}}{\rho\Omega}I_{1}\left(kr\right).
\end{equation}
With this in hand, through the kinematic condition equating the rate of change of the radius to the radial component of the fluid velocity at the boundary, $\partial_{t}R=v^{r}\left(R_{0}\right)$, the Rayleigh-Plateau dispersion relation emerges
\begin{equation}
	\Omega^{2}=\frac{\gamma I_{1}\left(kR_{0}\right)}{\rho R^{3}_{0} I_{0}\left(kR_{0}\right)}\left(k^{2}R^{2}_{0}-1\right)kR_{0},\label{eq:2.40}
\end{equation}
which is a hydrodynamic mode and exhibits dynamical instability in the long-wavelength regime $\lambda>2\pi R_{0}$.

Similar to the sound mode instability in viscous fluid, the Rayleigh-Plateau instability is likewise confined to the long-wavelength regime and leads to the breakdown of spatial translational invariance.
Differently, this hydrodynamic instability characterizes the properties of the fluid interface, leading to a topological transition of the interface.
In addition, the instability is directional, with its threshold determined solely by the geometric radius.
In section \ref{sec:Mi}, we demonstrate that the interface of Q-matter similarly suffers from this membrane instability, leading to a topological transition of the soliton.

\section{Sound mode instability in soliton field theory}\label{sec:Sis}
In this section, we investigate the stability of homogeneous configurations in soliton field theory.
Such configurations exhibit two distinct types of dispersion relation in the long-wavelength regime: one resembling that of a real scalar field mode, and the other analogous to a sound mode.
These give rise to tachyonic instability and sound mode instability, respectively.
The former drives the system to decay into other homogeneous configurations.
The latter, as the dynamical mechanism for soliton formation, breaks spatial translational invariance and decomposes the system into a coexistence state of the matter phase and the vacuum phase, resembling the phase separation process.

\subsection{Soliton fluid}
In the soliton field theory with Lagrangian density \eqref{eq:1.4}, the system is described by a Noether current associated with the $U(1)$ symmetry and by an energy-momentum tensor
\begin{subequations}
	\begin{align}
		J^{\mu}&=i\left(\psi^{*}\partial^{\mu}\psi-\psi\partial^{\mu}\psi^{*}\right),\\
		T^{\mu\nu}&=\partial^{\mu}\psi\partial^{\nu}\psi^{*}+\partial^{\mu}\psi^{*}\partial^{\nu}\psi-\eta^{\mu\nu}\mathcal{L}.
	\end{align}
\end{subequations}
With the definition
\begin{equation}
	\psi=e^{-i\vartheta}\phi,
\end{equation}
the four-velocity of the fluid associated with the complex scalar field, in the Eckart frame, is defined as
\begin{equation}
	u^{\mu}=\frac{J^{\mu}}{\sqrt{J^{\mu}J_{\mu}}}=\frac{\partial^{\mu}\vartheta}{\tilde{\vartheta}},
\end{equation}
with $\tilde{\vartheta}=\sqrt{\partial_{\mu}\vartheta\partial^{\mu}\vartheta}$.
Since the four-velocity is given by the gradient of the phase function, the system represents an irrotational fluid.
Furthermore, by decomposing the derivative of the profile function along the four-velocity
\begin{equation}
	\partial^{\mu}\phi=u^{\mu}D\phi+\Phi^{\mu},\label{eq:3.4}
\end{equation}
the associated hydrodynamic quantities can be expressed as
\begin{subequations}
	\begin{align}
		n&=2\tilde{\vartheta}\phi^{2},\\
		\epsilon&=\left(D\phi\right)^{2}+\tilde{\vartheta}^{2}\phi^{2}-\Phi_{\mu}\Phi^{\mu}+V,\\
		p+\Pi&=\left(D\phi\right)^{2}+\tilde{\vartheta}^{2}\phi^{2}+\frac{1}{3}\Phi_{\mu}\Phi^{\mu}-V,\\
		q^{\mu}&=2\Phi^{\mu}D\phi,\\
		\pi^{\mu\nu}&=2\Phi^{\mu}\Phi^{\nu}-\frac{2}{3}\Phi_{\alpha}\Phi^{\alpha}\Delta^{\mu\nu}.
	\end{align}
\end{subequations}
The fluid dynamics is governed by the conservation of the Noether current $\partial_{\mu}J^{\mu}=0$ and the energy-momentum tensor $\partial_{\mu}T^{\mu\nu}=0$, giving rise to
\begin{subequations}
	\begin{align}
		0&=\phi\partial_{\mu}\partial^{\mu}\vartheta+2\partial_{\mu}\phi\partial^{\mu}\vartheta,\label{eq:3.6a}\\
		0&=\partial_{\mu}\partial^{\mu}\phi-\phi\left(\tilde{\vartheta}^{2}- V'\right),\label{eq:3.6b}
	\end{align}\label{eq:3.6}
\end{subequations}
with $V'=dV/d|\psi|^{2}$.
Since the system involves only two independent field degrees of freedom, there exist correspondingly only two independent equations, which collectively constitute the Klein-Gordon equation $\partial_{\mu}\partial^{\mu}\psi+V'\psi=0$.

For a fluid at rest with four-velocity $u^{\mu}=\left(1,\vec{0}\right)$, the phase of the complex scalar field is a linear function of time, which we set as $\vartheta=\omega t,\omega>0$.
In this section, we focus on uniform configurations, which, as can be seen from the equation of motion, satisfy $\omega^{2}=V'$.
Equivalently, such states reside at the stationary points of the effective potential $U(\phi^{2})=V-\omega^{2}\phi^{2}$.
In this case, the nonvanishing physical quantities are
\begin{equation}
	n=2\omega\phi^{2},\quad \epsilon=\omega^{2}\phi^{2}+V,\quad p=\omega^{2}\phi^{2}-V,\label{eq:3.8}
\end{equation}
from which the equilibrium thermodynamic relation
\begin{equation}
	\epsilon+p=\omega n,
\end{equation}
can be extracted, indicating that the frequency $\omega$ can be interpreted as a chemical potential and that this system represents a cold matter with zero entropy.
In addition, defining two squared effective masses
\begin{equation}
	2m^{2}_{V}=\left(\phi^{2}V\right)'',\quad 2m^{2}_{U}=\left(\phi^{2}U\right)'',
\end{equation}
which are related by 
\begin{equation}
	m^{2}_{V}=m^{2}_{U}+\omega^{2},\label{eq:3.11}
\end{equation}
the squared speed of sound can be obtained as
\begin{equation}
	v^{2}_{s}=\frac{dp}{d\epsilon}=\frac{m^{2}_{U}}{m^{2}_{V}}.
\end{equation}
Therefore, the sign of the squared speed of sound is determined by the signs of the two squared effective masses.

\subsection{Dispersion relation}
To investigate the dynamical stability of the soliton fluid, similarly, a perturbation along the spatial $x$-direction is imposed on the equilibrium background
\begin{equation}
	\vartheta=\omega t+e^{-i\Omega t-ikx}\delta\vartheta,\quad \phi=\phi_{0}+e^{-i\Omega t-ikx}\delta\phi.
\end{equation}
Substituting it into the fluid dynamic equations \eqref{eq:3.6}, one obtains the following algebraic equations for the perturbations
\begin{equation}
	\begin{pmatrix}
		\left(-\Omega^{2}+k^{2}\right)\phi_{0}&-2i\omega\Omega\\
		2i\omega\Omega\phi_{0}&-\Omega^{2}+k^{2}+2\phi^{2}_{0}V''
	\end{pmatrix}\begin{pmatrix}
		\delta\vartheta\\
		\delta\phi
	\end{pmatrix}=0.
\end{equation}
The existence of nontrivial solutions requires that the determinant of the coefficient matrix vanish, namely
\begin{equation}
	0=(\Omega^{2}-k^{2})^{2}-4m_{V}^{2}\Omega^{2}+4m^{2}_{U}k^{2}.\label{eq:3.16}
\end{equation}
In the long-wavelength regime, this dispersion relation gives rise to a non-hydrodynamic mode and a hydrodynamic mode
\begin{subequations}
	\begin{align}
		\Omega^{2}_{m}&=4m^{2}_{V}+\left(1+\frac{\omega^{2}}{m^{2}_{V}}\right)k^{2}+o(k^{4}),\\
		\Omega^{2}_{h}&=v^{2}_{s}k^{2}+\frac{\omega^{4}}{4m^{6}_{V}}k^{4}+o(k^{6}).
	\end{align}
\end{subequations}
The non-hydrodynamic mode $\Omega_{m}$ resembles the dispersion relation of a real scalar field and, in the case of negative squared effective mass $m^{2}_{V}<0$, develops a dynamical instability, commonly referred to as tachyonic instability.
This instability characterizes the dynamical properties of an excited state located at a local maximum of the potential, which can decay into a ground state at a local minimum, thereby inducing a dynamical phase transition of matter, such as spontaneous scalarization in black hole backgrounds \cite{Doneva:2022ewd}.
Due to its non-hydrodynamic nature, this instability does not necessarily break spatial symmetries.
Interestingly, the soliton field resides at a stationary point of the effective potential $U$, whose nature as a local maximum or minimum is determined by the sign of the squared effective mass $2m_{U}^{2}=\phi^{2}U''$.
However, the tachyonic instability of soliton systems is governed by the sign of the squared effective mass $m^{2}_{V}$. 
Therefore, even at a local maximum $m^{2}_{U}<0$, the relation \eqref{eq:3.11} indicates that this is not sufficient to induce such an instability.
This exhibits a significant difference from the case of a real scalar field.
The other noteworthy outcome is the appearance of the hydrodynamic mode $\Omega_{h}$, which exhibits sound-wave propagation behavior.
However, such a dispersion relation is distinct from those of both ideal and viscous fluids.
The difference from the ideal fluid is obvious, manifested in the emergence of higher-order terms in the wave number.
For a stably propagating sound mode $v_{s}^{2}>0$, these terms lead to deviations from the ideal-fluid behavior in the short-wavelength regime.
When the sound mode instability, induced by a negative squared sound speed $v_{s}^{2}<0$, arises, these terms confine the instability to the long-wavelength regime, in a manner similar to that of a viscous fluid.
On the other hand, the distinction from a viscous fluid lies in the fact that the soliton field inherently lacks dissipation mechanisms.
This feature can be observed from the dispersion relation \eqref{eq:3.16}, which gives $\Omega^{2}=\Omega^{*2}$, indicating that stable modes are purely real and manifest as purely propagating behavior without decay, analogous to the case of an ideal fluid.
Another rationale is that an entropy-free fluid seems hardly susceptible to viscous effects, since their consequence would be to further reduce the entropy of the system, as indicated by equation \eqref{eq:2.17}.
In summary, the hydrodynamic mode $\Omega_{h}$ resembles that of an ideal fluid in the stable regime, while in the unstable regime it exhibits features analogous to those of a viscous fluid.

The dynamical properties of the system depend on whether it resides at a local minimum or a local maximum of the effective potential $U$.
In the former case, both squared effective masses are positive $m_{U}^{2}>0,m_{V}^{2}>0$, indicating dynamical stability of the system.
However, the issue arises in the latter case, where the squared effective mass $m_{U}^{2}<0$ becomes negative, leading to three possible scenarios.
The first is that the other squared effective mass remains positive $m_{V}^{2}>0$, in which case the system suffers from the sound mode instability but is spared from the tachyonic one.
The second is the opposite situation, corresponding to the negative squared effective mass $m_{V}^{2}<0$.
Interestingly, these two types of instabilities appear to be mutually exclusive.
The critical scenario is that of vanishing effective mass $m_{V}^{2}=0$, where the dispersion relation \eqref{eq:3.16} degenerates into 
\begin{equation}
	\Omega^{2}_{0}=\pm2\omega k+k^{2},\label{eq:3.18}
\end{equation}
with the negative branch signaling the onset of instability.
Such results indicate that the ground states are necessarily dynamically stable, whereas the excited states are inevitably unstable.
In addition, there exists a special case where the system resides at an inflection point, characterized by $m^{2}_{U}=0$ and $m^{2}_{V}=\omega^{2}$.
In this case, the dispersion relation \eqref{eq:3.16}, as a quadratic equation in $\Omega^{2}$, admits no negative roots, indicating that the system is dynamically stable.

\subsection{Sound mode instability and tachyonic instability}
Next, we turn to the soliton potential \eqref{eq:1.5} to discuss these scenarios in detail.
Since the ground states at local minima or inflection points have already been identified as dynamically stable, our focus will be on the excited states residing at local maxima.
Specifically, the scalar potential is chosen as
\begin{subequations}
	\begin{align}
		V_{1}&=m^{2}\phi^{2}+\lambda\phi^{4}+\kappa\phi^{6},\\
		V_{2}&=m^{2}\phi^{2+K}+\varkappa\phi^{6},
	\end{align}
\end{subequations}
with negative parameters $\lambda<0,K<0$ and a positive sextic term that dominates the potential for sufficiently large field value.
For both types of potentials, two of the parameters can be eliminated by rescaling the units of coordinates and energy,
\begin{subequations}
	\begin{align}
		\text{for }&V_{1}:\quad x:=mx,\quad \phi:=\frac{\phi\sqrt{|\lambda|}}{m},\quad\kappa:=\frac{m^{2}\kappa}{\lambda^{2}},\\
		\text{for }&V_{2}:\quad x:=\left(\frac{m^{2}}{\varkappa}\right)^{\frac{K}{2\left(4-K\right)}}mx,\quad \phi:=\left(\frac{\varkappa}{m^{2}}\right)^{\frac{1}{4-K}}\phi,
	\end{align}
\end{subequations}
which finally leads to
\begin{subequations}
	\begin{align}
		V_{1}&=\phi^{2}-\phi^{4}+\kappa\phi^{6},\\
		V_{2}&=\phi^{2+K}+\phi^{6}.
	\end{align}
\end{subequations}
Additionally, we further require $\kappa>1/4$ to ensure that the scalar potential $V_{1}$ possesses only a single non-degenerate vacuum $\phi=0$, and take $|K|\ll1$.

\begin{figure}
	\begin{center}
		\subfigure[]{\includegraphics[width=.49\linewidth]{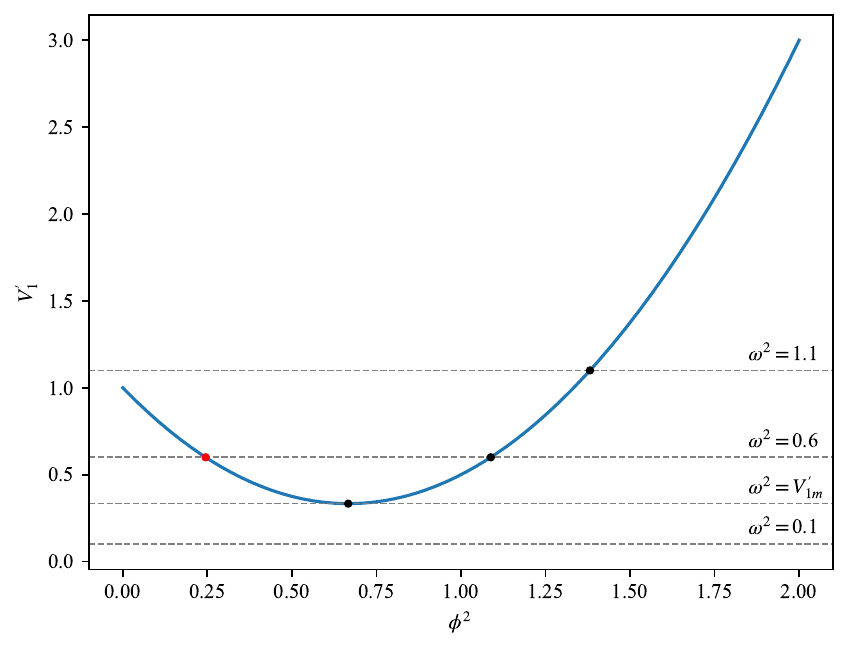}\label{fig:6}}
		\subfigure[]{\includegraphics[width=.49\linewidth]{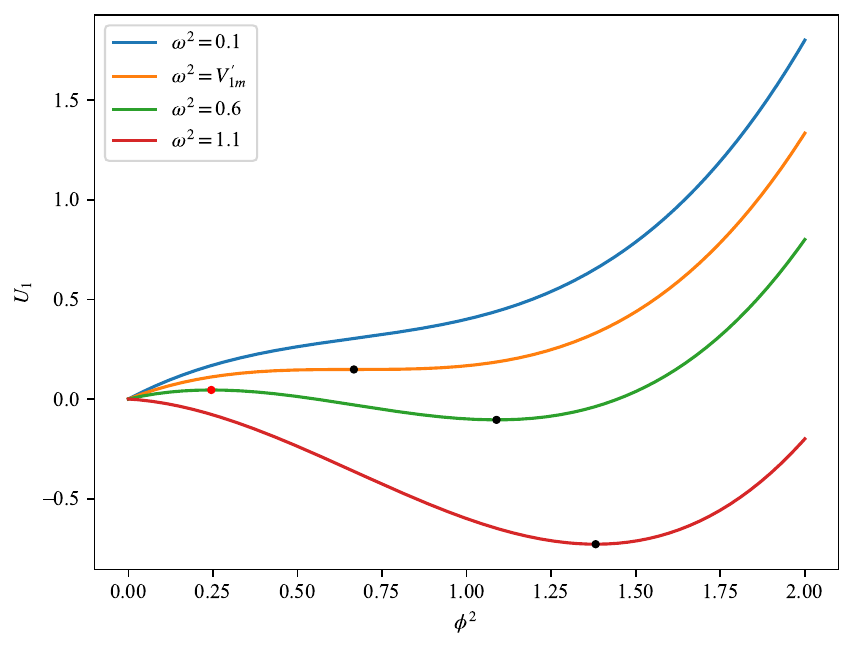}\label{fig:7}}
		\subfigure[]{\includegraphics[width=.49\linewidth]{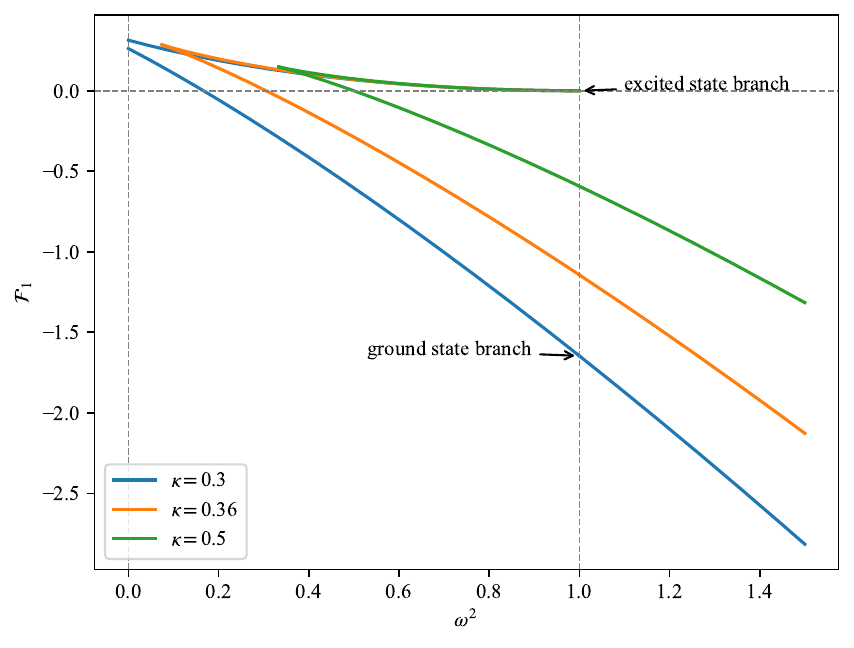}\label{fig:8}}
		\subfigure[]{\includegraphics[width=.49\linewidth]{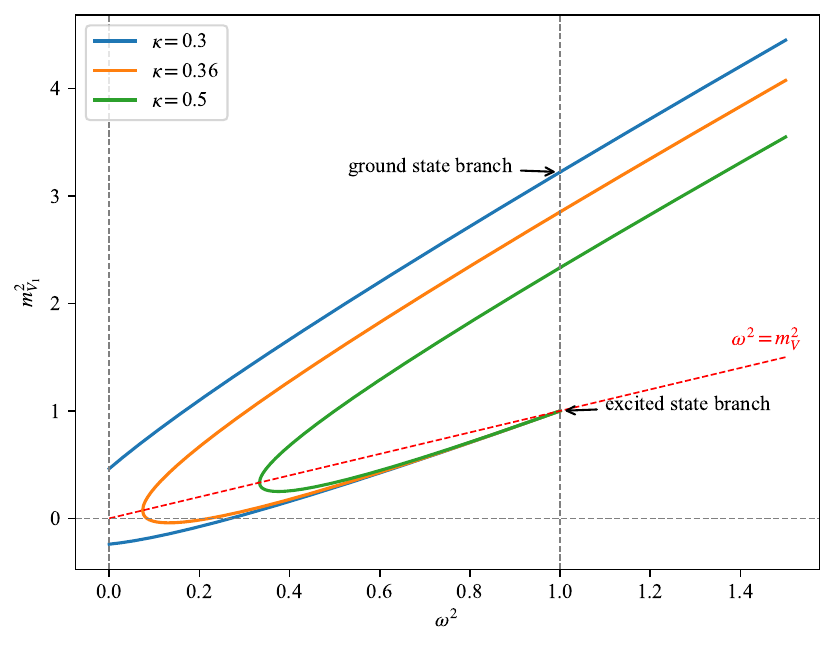}\label{fig:9}}
		\caption{The physical properties in the case of the scalar potential $V_{1}$.
			(a, b): The derivative of the scalar potential $V'$ and the effective potential $U$ as functions of the squared scalar field $\phi^{2}$, with parameter $\kappa=0.5$. 
			(c, d): The grand potential $\mathcal{F}$ and the squared effective mass $m^{2}_{V}$ as functions of the squared frequency $\omega^{2}$.}
		\label{fig:6-9}
	\end{center}
\end{figure}

Figure \ref{fig:6-9} illustrates the case of the scalar potential $V_{1}$.
For $\kappa>1/3$, the derivative of the scalar potential $V'_{1}$ possesses a positive minimum $V'_{1m}=1-\frac{1}{3\kappa}$, which divides the existence domain of the solutions into three regions.
Among them, non-vacuum homogeneous solutions appear in the region $\omega^{2}\geq V'_{1m}$.
When the squared frequency equals this minimum value, the effective potential $U_{1}$ develops an inflection point, signaling the emergence of non-vacuum solutions.
Subsequently, this inflection point splits into a local maximum at a small scalar field value and a local minimum at a large scalar field value.
Finally, as the squared frequency approaches and exceeds the squared scalar field mass $\omega^{2}\geq m^{2}=1$, the local maximum converges to the vacuum state and disappears from the domain, leaving a single ground-state non-vacuum solution.
On the other hand, for $1/4<\kappa<1/3$, since the minimum becomes negative $V'_{1m}<0$, non-vacuum solutions exist throughout the entire domain $\omega^{2}>0$, with excited states appearing in the region $0<\omega^{2}<1$.

For an entropy-free system with a chemical potential, the grand potential is defined as 
\begin{equation}
	\mathcal{F}=\epsilon-\omega n=-p,
\end{equation}
which is equal to the negative of the pressure.
Including the stable vacuum branch \footnote{The stability of the vacuum state is determined by the sign of the squared scalar field mass $m^{2}$, and in our case with a positive squared mass, the vacuum state is dynamically stable. However, the instability of the vacuum is in fact crucial for the generation of soliton field. During the inflation of the universe, when the negative mass-squared contribution from the Hubble parameter to the scalar field dominates, the tachyonic instability arises, leading to the exponential amplification of scalar fluctuations \cite{Dine:1995kz}.} with $\mathcal{F}=0$, figure \ref{fig:8} shows that the phase diagram of the fluid associated with the soliton field resembles that of a holographic fluid with a thermal first-order phase transition, as illustrated in figure \ref{fig:3}, displaying a characteristic swallowtail shape.
The ground-state branch, as discussed previously, is also dynamically stable.
Interestingly, the segment connected to the excited-state branch features a region of negative pressure.
This observation indicates that the occurrence of negative pressure does not necessarily entail dynamical instability.
Therefore, attributing the spatial instability of the Affleck–Dine field, which leads to Q-object formation, solely to negative pressure is not justified, even though the excited-state branch is always characterized by negative pressure.
The results shown in figure \ref{fig:9} suggest that the essence of this instability lies in the sound mode instability and the tachyonic instability.
As summarized in table \ref{table:1}, there are three distinct cases.
First, in the strong-coupling regime $3/8<\kappa$, the entire excited-state branch has positive squared effective mass $m^{2}_{V}>0$, indicating that the system exhibits a sound mode instability.
Second, in the intermediate-coupling regime $1/3<\kappa<3/8$, a small region $\frac{12\kappa-3-\sqrt{9-24\kappa}}{24\kappa}=\omega^{2}_{a}<\omega^{2}<\omega^{2}_{b}=\frac{12\kappa-3+\sqrt{9-24\kappa}}{24\kappa}$ where the squared effective mass becomes negative $m^{2}_{V}<0$ emerges on the excited-state branch, implying that the corresponding excited states develop a tachyonic instability while the sound mode remains stable.
In the other regions, the situation is reversed.
Finally, in the weak-coupling regime $1/4<\kappa<1/3$, the excited states in the low-frequency region $0<\omega^{2}<\omega^{2}_{b}$ exhibit a tachyonic instability, whereas those in the high-frequency region $\omega^{2}_{b}<\omega^{2}<1$ are affected by a sound mode instability.
In addition, at the interfaces between regions, $\omega^{2}=\omega^{2}_{a}$ or $\omega^{2}=\omega^{2}_{b}$, the excited states with vanishing effective mass $m^{2}_{V}=0$ are also dynamically unstable, exhibiting a dispersion relation \eqref{eq:3.18} that are distinct from those of both tachyonic instability and sound mode instability.

\begin{table}[h!]
	\centering
	\begin{tabular}{c|c|c|c}
		\hline
		Parameter range of $\kappa$& Region in $\omega^{2}$& The sign of $m^{2}_{V_{1}}$& Nature of instability\\
		\hline
		\hline
		$3/8<\kappa$&$(V'_{1m},1)$&$m^{2}_{V_{1}}>0$&Sound mode instability\\
		\hline
		\multirow{2}{*}{$1/3<\kappa<3/8$}&$(\omega^{2}_{a},\omega^{2}_{b})$&$m^{2}_{V_{1}}<0$& Tachyonic instability\\
		&$(V'_{1m},\omega^{2}_{a})\bigcup(\omega^{2}_{b},1)$&$m^{2}_{V_{1}}>0$&Sound mode instability\\
		\hline
		\multirow{2}{*}{$1/4<\kappa<1/3$}&$(0,\omega^{2}_{b})$&$m^{2}_{V_{1}}<0$&Tachyonic instability\\
		&$(\omega^{2}_{b},1)$&$m^{2}_{V_{1}}>0$&Sound mode instability\\
		\hline
	\end{tabular}
	\caption{Instability structure of the excited state branch in the case of the scalar potential $V_{1}$, with $V'_{1m}=1-\frac{1}{3\kappa}$, $\omega^{2}_{a}=\frac{12\kappa-3-\sqrt{9-24\kappa}}{24\kappa}$ and $\omega^{2}_{b}=\frac{12\kappa-3+\sqrt{9-24\kappa}}{24\kappa}$.}
	\label{table:1}
\end{table}

\begin{figure}
	\begin{center}
		\subfigure[]{\includegraphics[width=.49\linewidth]{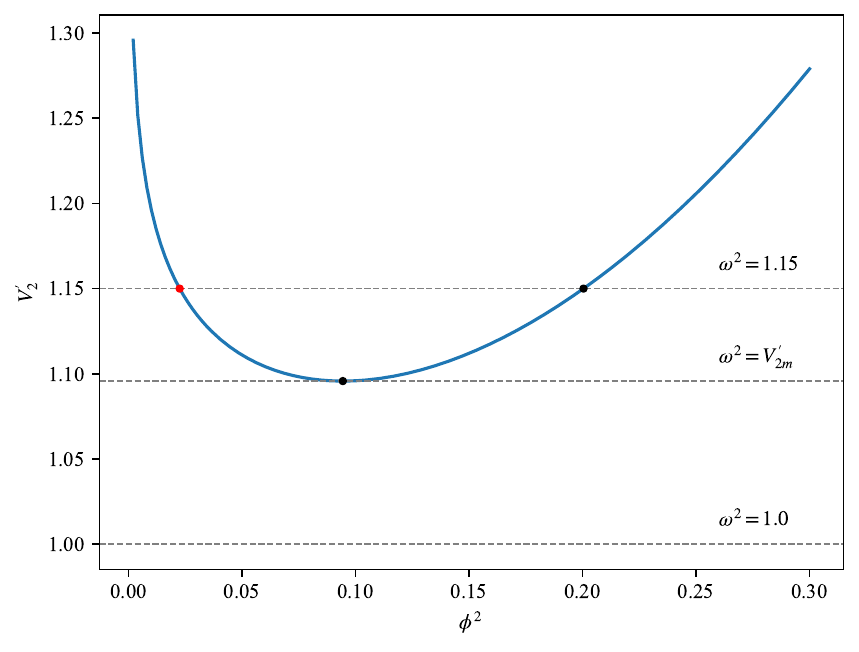}\label{fig:10}}
		\subfigure[]{\includegraphics[width=.49\linewidth]{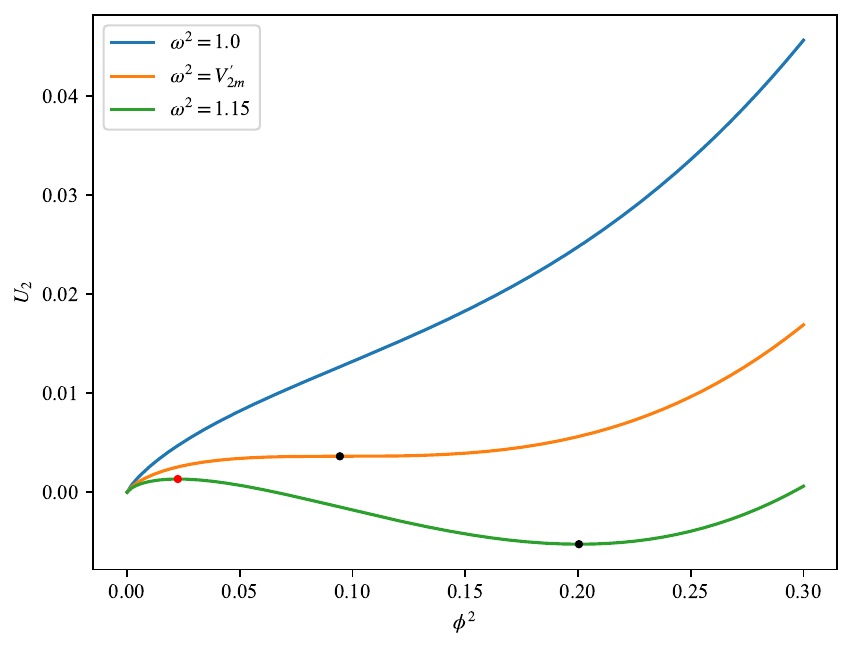}\label{fig:11}}
		\subfigure[]{\includegraphics[width=.49\linewidth]{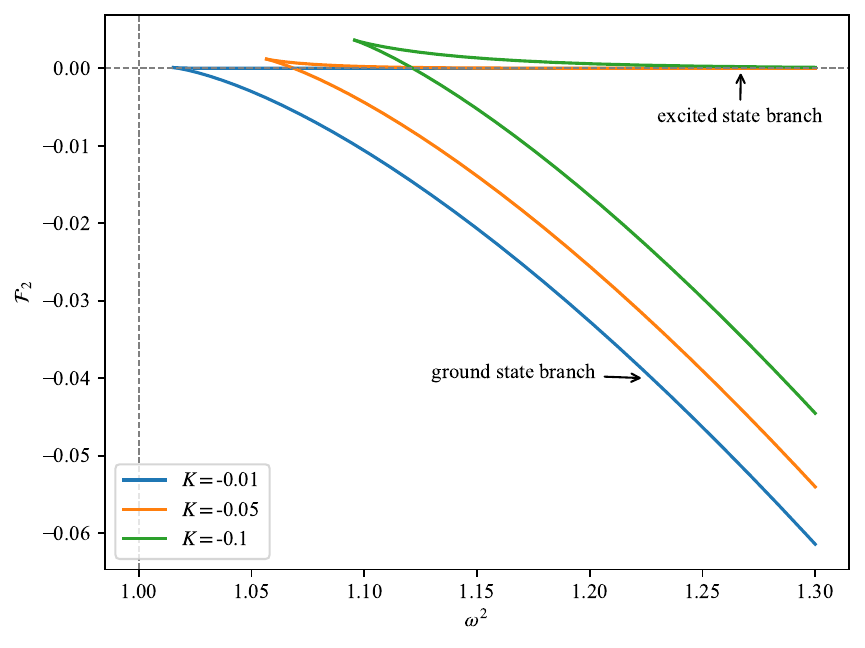}\label{fig:12}}
		\subfigure[]{\includegraphics[width=.49\linewidth]{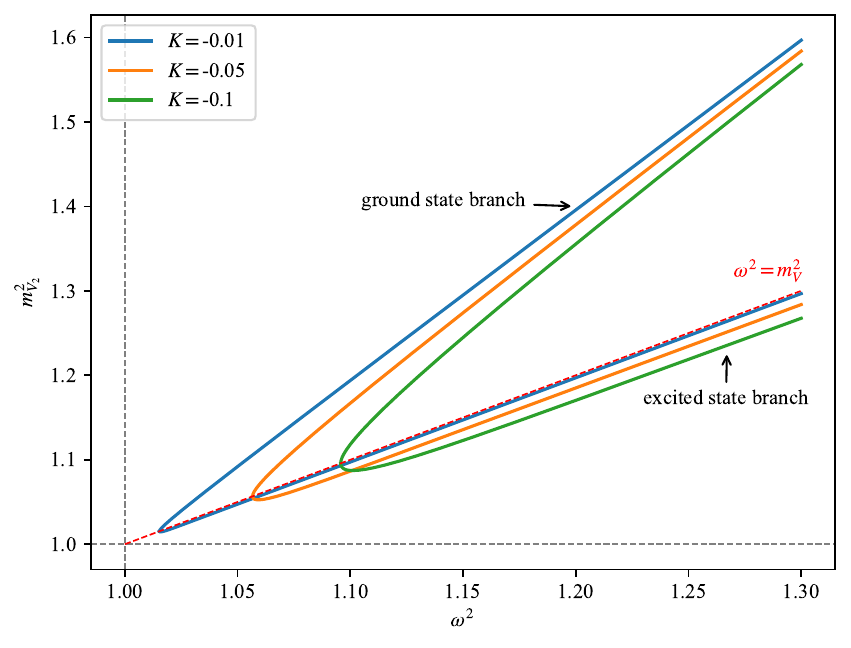}\label{fig:13}}
		\caption{The physical properties in the case of the scalar potential $V_{2}$.
			(a, b): The derivative of the scalar potential $V'$ and the effective potential $U$ as functions of the squared scalar field $\phi^{2}$, with parameter $K=-0.1$. 
			(c, d): The grand potential $\mathcal{F}$ and the squared effective mass $m^{2}_{V}$ as functions of the squared frequency $\omega^{2}$.}
		\label{fig:10-13}
	\end{center}
\end{figure}

The case of the scalar potential $V_{2}$ is shown in figure \ref{fig:10-13}, which closely resembles that of the scalar potential $V_{1}$ with several distinctive features.
First, the minimum of the derivative of the scalar potential always exceeds the squared scalar field mass $V'_{2m}>m^{2}=1$, which restrict the existence domain of non-vacuum solutions to the region $\omega^{2}\geq V'_{2m}>1$.
Moreover, the derivative of the scalar potential diverges near the vacuum $V'_{2}(0)\rightarrow\infty$, implying that the excited state associated with the local maximum of the effective potential $U_{2}$ necessarily coexists with the ground state at the local minimum throughout the entire existence domain.
Finally, a fundamental distinction from the case of $V_{1}$ lies in the fact that, in the scenario of $V_{2}$, the excited-state branch always carries a positive squared effective mass $m^{2}_{V}>0$ for any appropriate value of parameter $K$; consequently, the excited states are affected solely by the sound mode instability and remain completely free from the tachyonic instability.

\subsection{Phase separation and tachyonic decay}
Different categories of instabilities give rise to distinct dynamical phenomena.
For the sound mode instability driven by thermodynamic instability $dp/d\epsilon<0$, the excited states remain dynamically stable under homogeneous perturbations with $k=0$, but undergo a dynamical transition in response to long-wavelength perturbations.
This transition inevitably breaks spatial translational invariance and induces drastic changes in thermodynamic properties, leading to the emergence of multi-component coexistence states in which different spatial regions are occupied by distinct thermal phases connected by domain walls.
These newly generated components must be dynamically stable, and therefore thermodynamically stable $dp/d\epsilon>0$; otherwise, they will inevitably decay or transform into other stable configurations further.
In soliton field theory, the only possible candidates for such thermal phases are the vacuum phase and the ground-state non-vacuum phase, together forming a multi-component fluid analogous to that of holographic fluids.
In other words, the physical process of Affleck-Dine field condensation to form solitons can be intuitively understood as a phase separation mechanism: the homogeneous excited state partially transitions into the non-vacuum ground state in certain spatial region, while the remaining parts decay into the vacuum state, giving rise to particle-like soliton structures.

For multi-component coexistence states with a chemical potential, and neglecting interface effects, each phase must have the same grand potential in order to achieve dynamical equilibrium.
This requirement indicates that the grand potential of the non-vacuum phase inside the soliton structure vanishes, $\mathcal{F}=-p=0$, corresponding to the state represented by the intersection of the ground-state branch and the vacuum branch.
From this, one can derive the condition 
\begin{equation}
	\omega_{0}^{2}=\min\left(V/\phi^{2}\right),\label{eq:3.23}
\end{equation}
which precisely characterizes the thin-wall limit of the existence domain of the soliton.
This represents the ideal coexistence state in the absence of interface effects, constructed by stitching together homogeneous thermal phases of equal grand potential.
In realistic situations, however, interface effects, typically manifested as surface tension and size effects, lead to two deviations.
On the one hand, surface tension prevents the pressures of adjacent components from being equal.
Consequently, the grand potential of the internal phase of the soliton structure must be nonzero.
Moreover, since the surface tension at the interface points inward, the internal phase must possess a positive pressure, implying that this component should be composed of the non-vacuum ground state in the region $\omega^{2}_{0}<\omega^{2}$.
On the other hand, size effects cause the properties of each component in the coexistence state to deviate from those of the corresponding homogeneous thermal phases.
Therefore, when the interface size becomes non-negligible relative to the soliton scale, the internal component is bound to deviate more significantly from the characteristics of the non-vacuum ground state.
These two aspects will be illustrated in the following section.

Up to this point, we have implicitly considered the physical soliton configuration with vacuum outside and a matter field inside.
However, from the perspective of phase separation, the location of the formed thermal phase is not special, and the opposite scenario, vacuum inside and matter outside, is not prohibited; we refer to such structures as vacuum bubbles.
Since the direction of the surface tension depends solely on the curvature of the interface, it still points inward in this case.
Therefore, the pressure of the external matter phase can be concluded to be negative, indicating that it is composed of the non-vacuum ground state in the region $V'_{m}<\omega^{2}<\omega^{2}_{0}$.
For this scenario, the expected dynamical phenomenon is that the vacuum phase gradually grows in an excited state with negative pressure, eventually reaching equilibrium.
The external matter phase transitions into a non-vacuum ground state with negative pressure.
This process is analogous to the expansion of the early universe \cite{Guth:1980zm}.
This may be the significance of the existence and stability of the negative-pressure region on the ground-state branch.

The sound mode instability provides the dynamical mechanism for soliton formation from a homogeneous Affleck-Dine field.
Interestingly, in the case of scalar potential $V_{1}$, there exists a region where the excited states experience tachyonic instability while the sound mode remains stable, enabling dynamical transitions even under homogeneous perturbations with $k=0$.
Since this transition does not necessarily break the spatial translational invariance, the final state is a homogeneous phase rather than a soliton structure.
On the other hand, due to the conservation of the particle number, it is also not the vacuum phase.
Intuitively, it should be the dynamically stable non-vacuum ground states; however, the excited states with sound mode instability are also dynamically stable under homogeneous perturbations, making them equally viable candidates for the final state.

\begin{figure}
	\begin{center}
		\subfigure[]{\includegraphics[width=.49\linewidth]{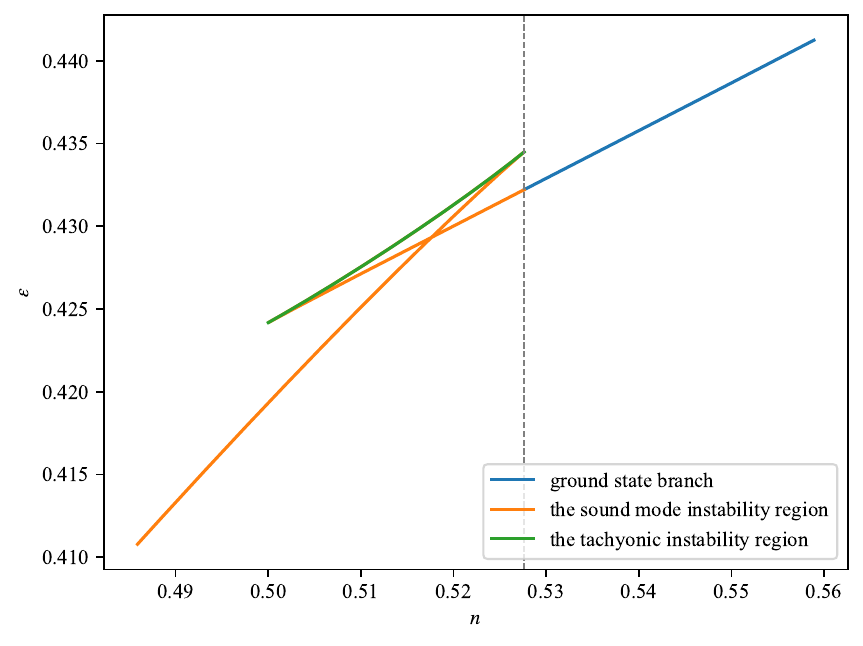}\label{fig:14}}
		\subfigure[]{\includegraphics[width=.49\linewidth]{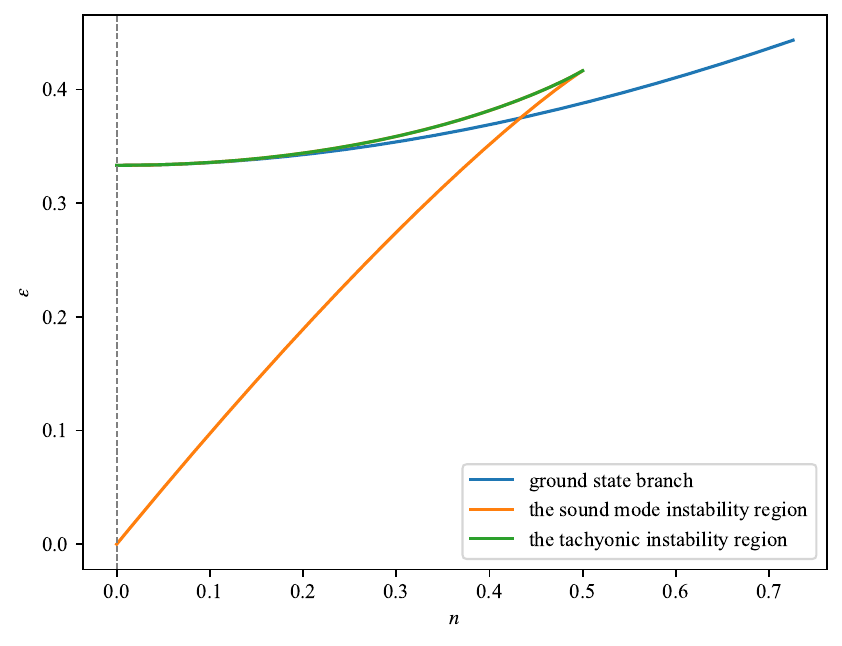}\label{fig:15}}
		\caption{In the case of scalar potential $V_{1}$, the energy density of non-vacuum states as a function of particle number density with parameter $\kappa=\kappa_{c}\approx0.36338998$ (a) and $\kappa=1/3$ (b).}
		\label{fig:14-15}
	\end{center}
\end{figure}

\begin{figure}
	\centering\includegraphics[width=.9\linewidth]{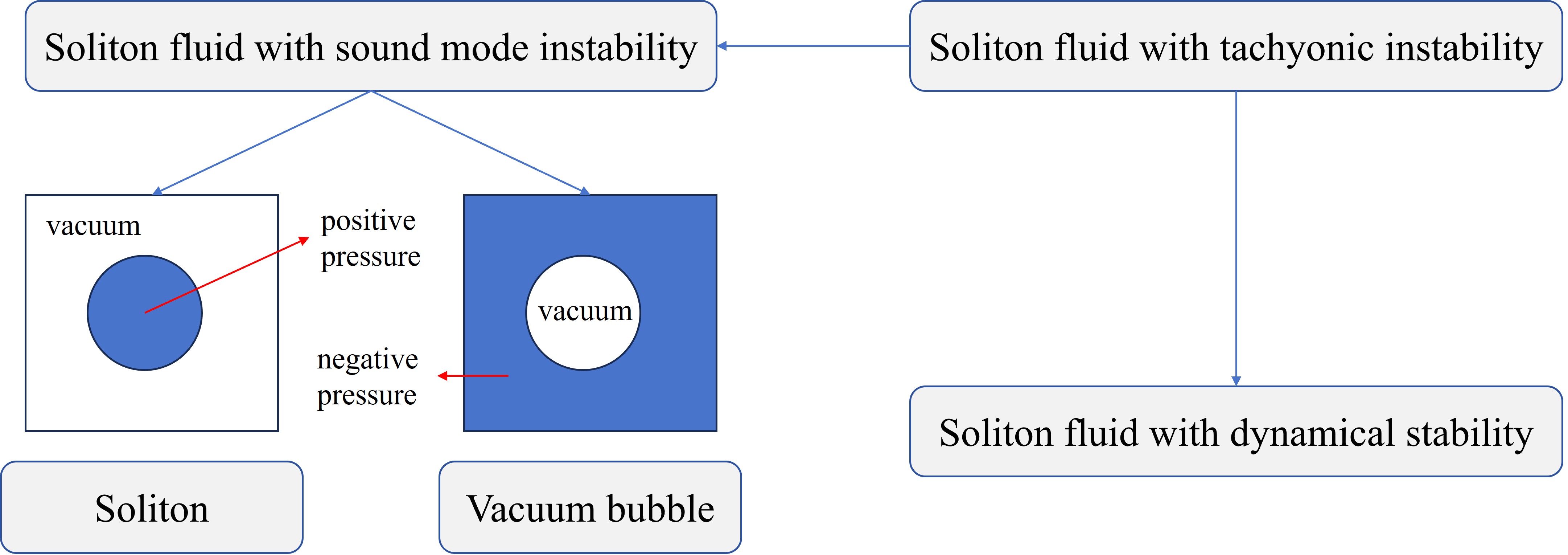}
	\caption{Possible evolution pathways of soliton fluids.}
	\label{fig:22}
\end{figure}

The answer to this issue is illustrated in figure \ref{fig:14-15}, which shows the change of energy density with particle number density.
For a system with fixed particle number, the excited states with tachyonic instability possess the highest energy and can therefore decay into lower-energy states.
Interestingly, there are two candidates for such lower-energy states, one of which is necessarily an excited state with sound mode instability.
Hence, the dynamical pathway from a tachyonically unstable excited state to an excited state with sound mode instability is guaranteed to exist.
The identity of the second candidate depends on the parameter $\kappa$ and the particle number of the system, with a critical value $\kappa_{c}$ dividing the scenarios into three cases.
First, for $\kappa_{c}<\kappa<3/8$, the second candidate is also an excited state with sound mode instability.
Second, for $1/3<\kappa<\kappa_{c}$, the intersection of the ground-state and excited-state branches (represented by the gray dashed line in the figure) enters the tachyonically unstable region, implying that the second candidate could be either an excited state with sound mode instability or a non-vacuum ground state, depending on the particle number of the system.
Furthermore, as $\kappa$ approaches $1/3$, the region where the non-vacuum ground state serves as the candidate gradually expands.
Until, for $1/4<\kappa<1/3$, the region with sound mode instability connected to the ground-state branch disappears completely, indicating that in this case the second candidate can only be the non-vacuum ground state.
In summary, as shown in figure \ref{fig:22}, the dynamical evolution originating from the excited states can proceed along three distinct pathways:
\begin{itemize}
	\item excited state with sound mode instability $\rightarrow$ soliton or vacuum bubble,
	\item excited state with tachyonic instability $\rightarrow$ non-vacuum ground state,
	\item excited state with tachyonic instability $\rightarrow$ excited state with sound mode instability $\rightarrow$ soliton or vacuum bubble.
\end{itemize} 
Such results indicate that the well-known pathway, in which a homogeneous Affleck-Dine field fragments into solitons \cite{Kasuya:1999wu,Kasuya:2000wx}, is not the only possible dynamical evolution route.
Within certain parameter regimes, the system may instead undergo a tachyonic decay directly into a homogeneous non-vacuum ground state, thereby preserving spatial symmetry.
In addition, in other parameter ranges, the Affleck-Dine field can tachyonically decay into another Affleck-Dine configuration with sound mode instability, which then serves as a dynamical intermediate state and subsequently fragments into solitons or vacuum bubbles.
The latter two evolution scenarios require further verification through numerical simulations.


\section{Membrane instability in soliton field theory}\label{sec:Mi}
In this section, by analogy with a water column, we investigate the stability of a class of cylindrically symmetric configurations in soliton field theory, referred to as Q-strings.
Similarly, these solitons exhibit dynamical instability to perturbations along the cylindrical direction in the long-wavelength regime.
When the surface energy density is much smaller than the bulk energy density, the dispersion relation of this instability degenerates to the Rayleigh-Plateau dispersion relation in the thin-wall limit.
Such an instability can reduce the surface area of the soliton, thereby triggering a topological transition from cylindrical symmetry to spherical symmetry, resulting in the formation of Q-balls.

\subsection{Q-strings}
The soliton formation process through the Affleck-Dine mechanism can be interpreted as a phase separation phenomenon triggered by the sound mode instability, which inevitably leads to the emergence of interfaces separating distinct thermal phases.
In physics, the topological structure and dynamical properties of such interfaces are also an intriguing topic, constituting an important part of interface physics.
In general, within a phase separation mechanism, the topology of the emergent interfaces is determined by the symmetry of the applied perturbations.
Therefore, the interfaces of phase-separated states can naturally exhibit planar symmetry, cylindrical symmetry, spherical symmetry.
For solitons, the stationary equation governing the profile function of the system with these topological structures can be expressed in the following unified form
\begin{equation}
	\frac{d^{2}}{dr^{2}}\phi=-F-\frac{d}{d\phi}\widetilde{U},\label{eq:4.1}
\end{equation}
with $\widetilde{U}(\phi)=-\frac{1}{2}U(\phi)$.
Such an equation can be intuitively interpreted as the Newtonian equation of a unit-mass classical particle moving in a potential $\widetilde{U}$, subject to a friction force $F$, with position $\phi$ and time $r$.
The specific form of the friction force depends on the symmetry of the profile function, given explicitly as $F=\frac{D}{r}\frac{d\phi}{dr}$, where $D=0,1,2$ represent the cases of planar symmetry, cylindrical symmetry, and spherical symmetry, respectively.
In this work, drawing an analogy with a water column, we focus on the case with cylindrical symmetry, where $F=\frac{1}{r}\frac{d\phi}{dr}$.
Under this consideration, a soliton solution corresponds to a trajectory that starts from position $\phi=\phi(0)$ at time $r=0$ and terminates at the origin $\phi=0$ after infinite time $r\rightarrow\infty$.
Due to the effect of friction, the mechanical energy $E_{m}=\frac{1}{2}\left(\frac{d\phi}{dr}\right)^{2}+\widetilde{U}$ of the particle decreases continuously along the motion.
Therefore, a necessary condition for the trajectory to reach the endpoint is that the mechanical energy at the initial position is greater than that at the end point.
By performing a Taylor expansion of the kinematic equation \eqref{eq:4.1} at the initial and final positions, one can find the kinetic energy of the particle vanishes at both points $\frac{d\phi}{dr}(r=0)=\frac{d\phi}{dr}(r\rightarrow\infty)=0$.
Consequently, the existence condition of the trajectory depends solely on the potential energy $\widetilde{U}(\phi(0))>\widetilde{U}(0)=0$, which yields $\omega^{2}>\min\left(V/\phi^{2}\right)$.
This result is consistent with the soliton existence domain \eqref{eq:3.23} obtained from the viewpoint of phase separation.

\begin{figure}
	\begin{center}
		\subfigure[]{\includegraphics[width=.49\linewidth]{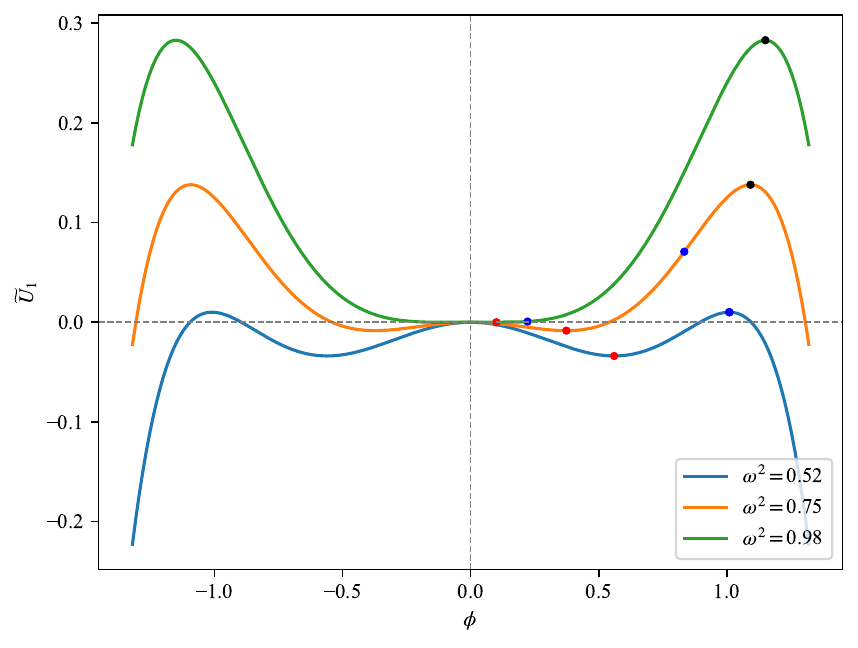}\label{fig:16}}
		\subfigure[]{\includegraphics[width=.49\linewidth]{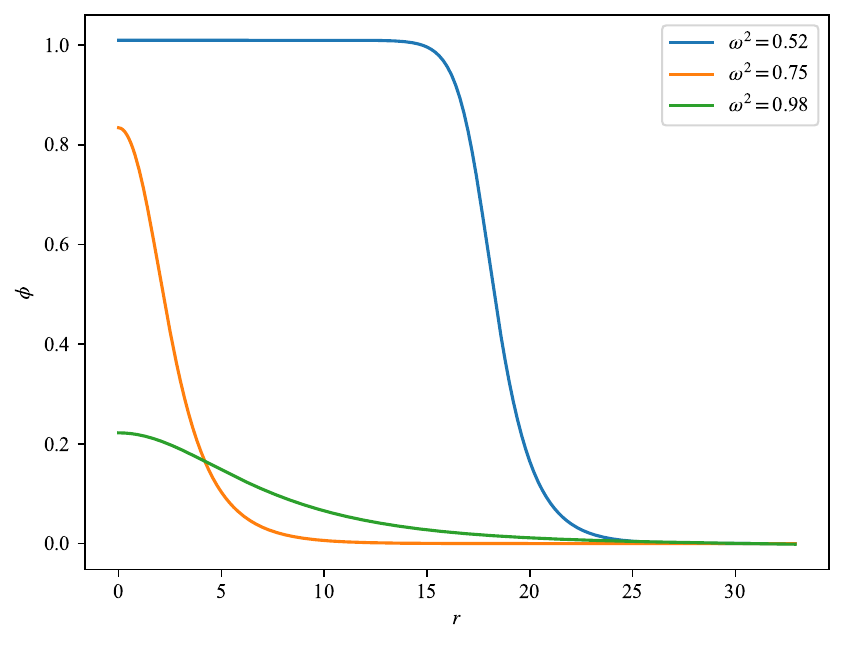}\label{fig:17}}
		\subfigure[]{\includegraphics[width=.49\linewidth]{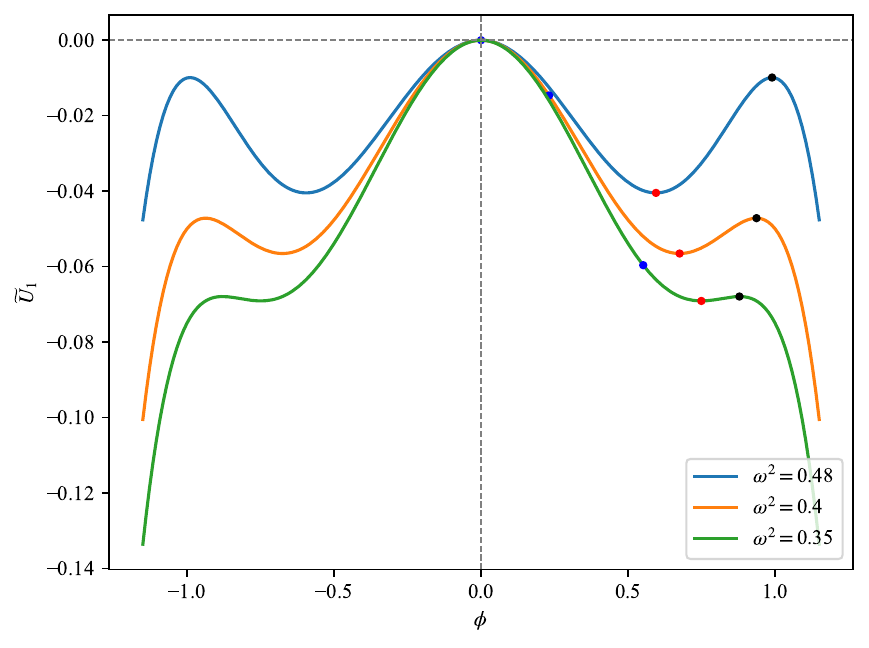}\label{fig:18}}
		\subfigure[]{\includegraphics[width=.49\linewidth]{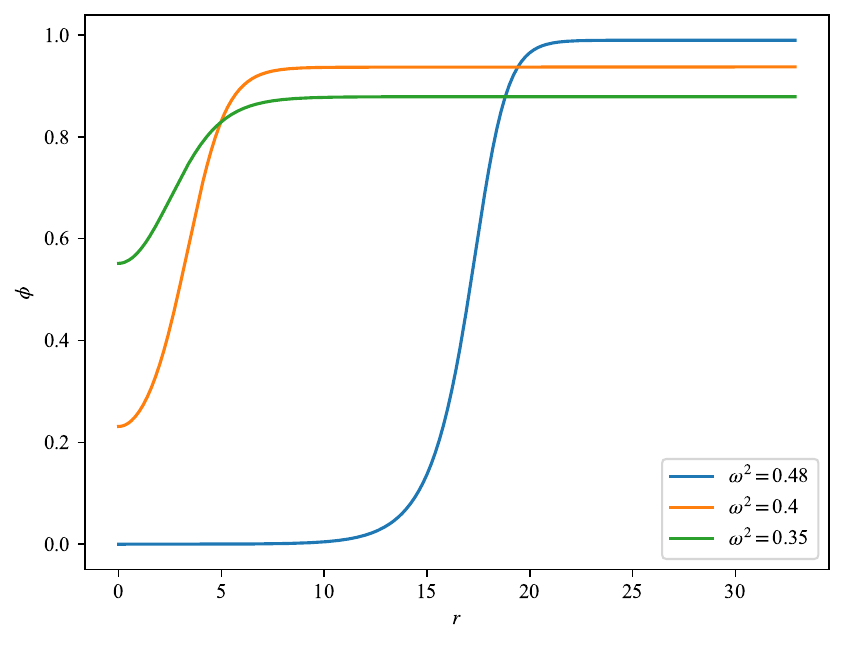}\label{fig:19}}
		\caption{In the case of scalar potential $V_{1}$, the schematic diagrams of solitons and vacuum bubbles. (a, c): The particle potential $\widetilde{U}$ as a function of the scalar field $\phi$ with parameter $\kappa=0.5$. The upper and lower panels correspond to the cases $\omega^{2}>\omega^{2}_{0}$ and $\omega^{2}<\omega^{2}_{0}$, representing the regions where solitons and vacuum bubbles exist, respectively. The red and black dots correspond to the excited and ground states in figure \ref{fig:7}, and the blue dot marks the initial position of the particle trajectory, representing the internal matter field of the soliton and the vacuum bubble. (b, d): The configurations of solitons and vacuum bubbles.}
		\label{fig:16-19}
	\end{center}
\end{figure}

Taking the case of scalar potential $V_{1}$ with parameter $\kappa=0.5$ as an example, figure \ref{fig:16-19} illustrates the schematic diagrams of the particle trajectories and the geometric configurations of solitons and vacuum bubbles.
As shown in figure \ref{fig:16}, for $\omega^{2}_{0}<\omega^{2}$, between the excited state (red point) and the non-vacuum ground state (black point), there exists a critical value of the scalar field (blue point) \footnote{Excluding the case of excited solitons whose profile functions possess nodes \cite{Mai:2012cx}.}.
For initial positions smaller than this threshold, the particle trajectory converges oscillatory to the positive excited state, whereas for those larger than the threshold, it converges oscillatory to the negative excited state.
At the critical value, the trajectory monotonically decays to the origin, corresponding to a soliton configuration, as shown in figure \ref{fig:17}.
As the frequency approaches the boundary of the soliton existence domain $\omega^{2}\rightarrow\omega^{2}_{0}$, the relation $\mathcal{F}=-2\widetilde{U}$ ensures that the particle potential at the non-vacuum ground state necessarily tends to zero.
In this limit, the initial position of the particle approaches the location of the non-vacuum ground state, indicating that the properties of the soliton interior increasingly resemble those of the homogeneous non-vacuum ground state.
Due to the flatness near the extremum, the particle remains longer at its initial position, implying that the interface size is negligible compared to the soliton scale;
this limit is thus referred to as the thin-wall limit.
Away from this limit, size effects of the interface gradually becomes dominant, and the properties of the soliton interior increasingly deviate from those of the non-vacuum ground state, as reflected by the displacement of the initial position of the particle from that of the non-vacuum ground state.
Once the frequency exceeds the scalar field mass $m^{2}<\omega^{2}$, the origin turns into a local minimum, prohibiting the existence of solitons.
Since the interface of the soliton structure is very smooth in this regime, it is referred to as the thick-wall limit \footnote{This limit does not arise in the case of scalar potential $V_{2}$.}.
The above phenomenon suggests that the soliton structure can be regarded as a phase-separated configuration, deviating from the ideal coexistence state due to interface effects, as discussed previously.
Moreover, from the perspective of phase separation, configurations with an internal vacuum and an external matter phase arise naturally.
As shown in figure \ref{fig:18}, for $\omega^{2}<\omega^{2}_{0}$, the particle potential at the non-vacuum ground state is negative, and thus there is no trajectory corresponding to a soliton.
However, there exists a critical value of the scalar field between the vacuum and the excited states, such that a particle trajectory starting from this value terminates at the non-vacuum ground state, as illustrated in figure \ref{fig:19}.
This is precisely a structure opposite to a physical soliton, with the difference that the external matter phase exhibits negative pressure.
Away from the thin-wall limit, the vacuum bubble exhibits behaviors similar to those of a soliton, with the internal phase gradually deviating from the vacuum state due to interface effects.
Until $\omega^{2}<V'_{1m}$, the vacuum bubble structure disappears.

For a static soliton system, the dependence of the profile function on the spatial coordinates implies that the spatial part of the decomposition \eqref{eq:3.4}, $\Phi^{\mu}$, is necessarily nonzero.
Since this term involves only spatial derivatives of the profile function, it contributes solely at the interface and can thus be regarded as a surface term.
Accordingly, the energy-momentum tensor can be separated into internal and surface parts
\begin{equation}
	T^{\mu\nu}=T_{\text{interior}}^{\mu\nu}+T_{\text{surface}}^{\mu\nu}.
\end{equation}
Among them, the nonvanishing physical quantities contained in the former coincide with those of the homogeneous soliton fluid given in \eqref{eq:3.8}
\begin{equation}
	T_{\text{interior}}^{\mu\nu}=\epsilon u^{\mu}u^{\nu}-p\Delta^{\mu\nu},
\end{equation}
accompanied by a particle number density $n$.
Since the soliton interface can be defined by $\phi\left(\vec{x}\right)=$constant, the surface term $\Phi^{\mu}=\nabla^{\mu}\phi$ is a space-like vector normal to the surface.
After normalization $\Phi^{\mu}=\widetilde{\Phi}\xi^{\mu}$ with $\widetilde{\Phi}=\sqrt{-\Phi^{\mu}\Phi_{\mu}}$ denoting the modulus of the vector and $\xi^{\mu}$ being the unit space-like normal vector to the interface, the surface contribution to the energy-momentum tensor can be expressed as
\begin{equation}
	T_{\text{surface}}^{\mu\nu}=\varsigma \left(\frac{1}{2}u^{\mu}u^{\nu}+\widetilde{\Delta}^{\mu\nu}+\frac{1}{6}\Delta^{\mu\nu}\right),
\end{equation}
where $\widetilde{\Delta}^{\mu\nu}=\xi^{\mu}\xi^{\nu}+\Delta^{\mu\nu}/3$ represents the traceless part of the induced spatial metric and $\varsigma=2\widetilde{\Phi}^{2}=2\left(d\phi/dr\right)^{2}$ characterizes the distribution of shear force.
The surface tension is defined as the projection onto the radial direction of the integral of the shear force along the interface normal vector
\begin{equation}
	\gamma=\int dr\xi_{r}\varsigma=\int dr\varsigma.
\end{equation}
To be consistent with the convention in figure \ref{fig:5}, we take the normal vector to point inward, denoted by $\xi_{r}=-\xi^{r}>0$.
Using the shear force as a measure, one can further define the effective radius of a soliton
\begin{equation}
	R=\frac{\int dr\left(r\varsigma\right)}{\int dr\varsigma}.\label{eq:4.6}
\end{equation}
With this definition, the surface energy of the Q-string of length $L$ is given by
\begin{equation}
	E_{\text{surface}}=\int dx^{3}\varsigma=2\pi R L\gamma.\label{eq:4.7}
\end{equation}
Therefore, surface tension can serve as a characterization of the surface energy density.

\subsection{Membrane instability}
By numerically solving the perturbation equations, we have demonstrated in the previous work \cite{Chen:2024axd} that Q-strings exhibit a long-wavelength hydrodynamic instability along the cylindrical direction.
In the thin-wall regime, the dispersion relation of such an instability approaches that in the Rayleigh-Plateau model \eqref{eq:2.40}, indicating the hydrodynamic nature of the soliton interface.
Building upon this result, we proceed to perform a more detailed analysis of the characteristics of membrane instability of Q-strings in the thin-wall limit.

Before that, let us reexamine the stationary equation \eqref{eq:4.1} that determines the soliton configuration, which is interpreted in the previous subsection as the Newtonian equation for the motion of a classical particle.
Equivalently, it can be expressed in the form of energy conservation
\begin{equation}
	\left[\frac{1}{2}\left(\frac{d\phi}{dr}\right)^{2}+\widetilde{U}\right]^{\infty}_{0}=-\int_{0}^{\infty}dr\left(F\frac{d\phi}{dr}\right),
\end{equation}
which characterizes the balance that the loss of mechanical energy equals the work performed by the friction force.
Due to the relations $\frac{d\phi}{dr}\left(r=0\right)=\frac{d\phi}{dr}\left(r\rightarrow0\right)=0$, $\widetilde{U}=p/2$, and $F=\frac{D}{r}\frac{d\phi}{dr}$, the above equation can be further reformulated into the form
\begin{equation}
	p\left(0\right)-p\left(\infty\right)=D\int_{0}^{\infty}dr\frac{\varsigma}{r},\label{eq:4.9}
\end{equation}
which characterizes the balance between the pressure difference inside and outside the soliton and the interface effects.
Defining the right-hand side of the equation as the interface pressure $p=D\int_{0}^{\infty}dr\frac{\varsigma}{r}$, elegantly, this structure is precisely analogous to the Young-Laplace equation \eqref{eq:2.30} that characterizes the pressure induced by surface tension in a liquid membrane.
In the zero-thickness interface limit, the interface pressure can be approximated as $p=\gamma\frac{D}{R}$, with $\frac{D}{R}$ representing the principal curvature of the interface, corresponding respectively to a plane $(D=0)$, a cylinder $(D=1)$, and a sphere $(D=2)$, indicating a reduction to the Young-Laplace equation.
Such results indicate that the soliton interface resembles a liquid membrane with surface tension, thereby demonstrating its hydrodynamic nature.

We are now in a position to investigate the membrane instability of Q-strings in the thin-wall limit.
For simplicity, we assume that the soliton interface has zero thickness, in which case the profile function can be expressed as
\begin{equation}
	\phi=\phi_{\text{in}} X\left(R-r\right),
\end{equation}
where $\phi_{\text{in}}$ is the internal field value, $R$ is the radius, and $X(x)=\left\lbrace 0,x<0;\frac{1}{2},x=0;1,x>0\right\rbrace $ denotes the step function.
Under this approximation, the surface tension is given by
\begin{equation}
	\gamma=2\phi^{2}_{\text{in}}\int_{0}^{\infty}dr\Xi^{2}\left(R-r\right),
\end{equation}
where $\Xi\left(x\right)=\frac{d}{dx}X\left(x\right)$ denotes the delta function.
Although this is not rigorously well-defined, it does not hinder further exploration of the problem.
Such divergent behavior can be remedied by taking into account a finite interface thickness \cite{Heeck:2020bau}.
In addition, we further assume that the internal matter of the soliton is incompressible, meaning that the energy density and particle number density remain unchanged in time, equivalently $\partial_{t}\phi_{\text{in}}=\partial_{t}\tilde{\vartheta}=0$.
Such an assumption leads to two consequences.
First, the variation of the soliton's profile function manifests only in its radius
\begin{equation}
	\phi=\phi_{\text{in}}X\left(R_{0}-r\right)+\phi_{\text{in}}\Xi\left(R_{0}-r\right)\delta R+o\left(\delta^{2}\right).
\end{equation}
From the definition \eqref{eq:4.6}, the perturbation of the radius can be expressed as $R=R_{0}+\delta R$.
Second, the second-order time derivative of the variation of the phase function is of higher-order smallness.
Considering the perturbation of the phase function
\begin{equation}
	\vartheta=\omega t+\delta\vartheta+o(\delta^{2}),
\end{equation}
one obtains
\begin{equation}
	\tilde{\vartheta}=\sqrt{\partial_{\mu}\vartheta\partial^{\mu}\vartheta}=\omega+\partial_{t}\delta\vartheta+o(\delta^{2}),
\end{equation}
so incompressibility indicates
\begin{equation}
	\partial^{2}_{t}\delta\vartheta\sim o(\delta^{2}).
\end{equation}
With these assumptions, the field equations \eqref{eq:3.6} degenerate into a simple form that governs the distribution of internal velocity potential and the dynamics of the interface.

\begin{figure}
	\centering\includegraphics[width=.49\linewidth]{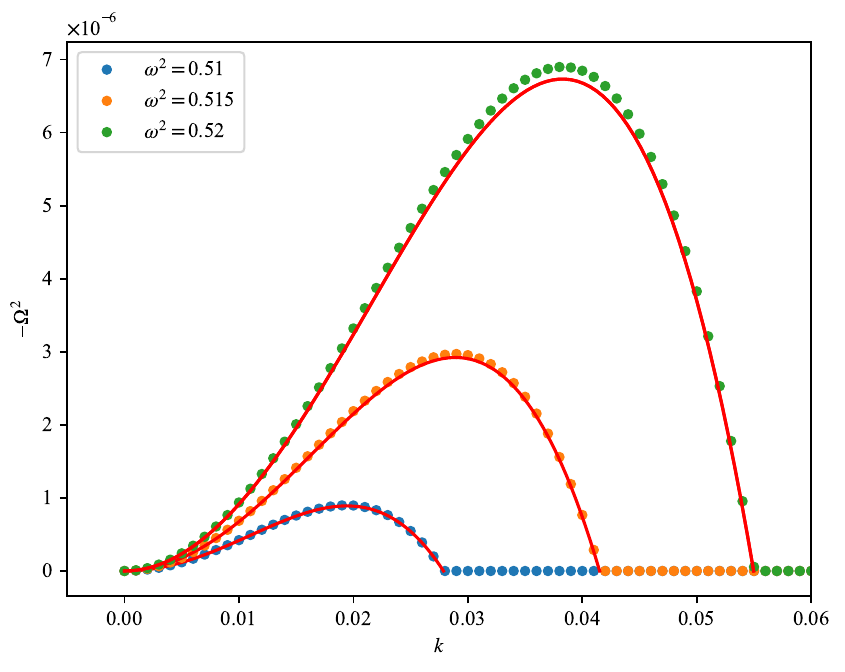}
	\caption{Dispersion relation of the membrane instability of Q-strings. Dots represent the numerical results obtained by solving the linearized Klein-Gordon equation without approximation using the spectral decomposition method, while the line denotes the analytical expression \eqref{eq:4.24}. Dots of different colors correspond to different oscillation frequencies. The scalar potential is chosen as $V_{1}$ with parameter $\kappa=0.5$, in which case the thin-wall limit is $\omega^{2}_{0}=0.5$.}
	\label{fig:20}
\end{figure}

To investigate the membrane instability of the Q-string, we impose a perturbation with wavelength $\lambda=2\pi/k$ along the cylindrical direction
\begin{equation}
	\vartheta=\omega t + e^{-i\Omega t-ikz}\delta\vartheta\left(r\right), \quad R=R_{0} + e^{-i\Omega t-ikz}\delta R,\quad p=p_{0}+e^{-i\Omega t-ikz}\delta p\left(r\right).
\end{equation}
Accordingly, the first-order perturbation of the Noether current conservation equation \eqref{eq:3.6a} takes the following form
\begin{equation}
	\left(r^{2}\partial^{2}_{r}\delta\vartheta+r\partial_{r}\delta\vartheta-k^{2}r^{2}\delta\vartheta\right)X-2r^{2}\left(\partial_{r}\delta\vartheta-i\omega\Omega\delta R\right)\Xi=0.
\end{equation}
The first term describes the distribution of the velocity potential inside the soliton, which satisfies a modified Bessel equation with solution
\begin{equation}
	\delta\vartheta\left(r\right)=CI_{0}\left(kr\right).
\end{equation}
Owing to the regularity at the origin, only the Bessel function of the first kind $I\left(r\right)$ is retained, while the second kind $K\left(r\right)$ is discarded.
The second term provides the relation between the interface velocity and the velocity potential
\begin{equation}
	i\omega\Omega\delta R=\partial_{r}\delta\vartheta\left(R_{0}\right)=kCI_{1}\left(kR_{0}\right).\label{eq:4.19}
\end{equation}
The perturbation of the velocity potential naturally induces a pressure fluctuation inside the soliton, whose distribution is given by
\begin{equation}
	\delta p_{\tilde{\vartheta}}\left(r\right) =\frac{\partial p}{\partial\tilde{\vartheta}}\delta\tilde{\vartheta}=-in_{0}\Omega CI_{0}\left(kr\right)X^{2}(R_{0}-r),
\end{equation}
with the particle number density $n_{0}=2\omega\phi^{2}_{\text{in}}$. 
The pressure fluctuation at the soliton interface is governed by the energy-momentum tensor conservation equation \eqref{eq:3.6b}, which characterizes the relation between the interface acceleration and the pressure variation.
Similar to the stationary equation \eqref{eq:4.9}, the energy-momentum tensor conservation equation can be expressed in terms of the pressure gradient, thereby taking the structure of the Young-Laplace equation
\begin{equation}
	\frac{dp}{dr}=2\partial_{\mu}\partial^{\mu}\phi\frac{d\phi}{dr}+n\frac{d\tilde{\vartheta}}{dr},
\end{equation}
with $n=n_{0}X^{2}\left(R-r\right)$, whose first-order perturbation is given by
\begin{equation}
	\frac{d}{dr}\left(\delta p-\delta p_{\tilde{\vartheta}}\right)\left(r\right)=\left[\varsigma\left(\Omega^{2}+\frac{1}{r^{2}}- k^{2}\right)+\frac{d}{dr}\left(\frac{\varsigma}{r}\right)+\frac{d^{2}}{dr^{2}}\left(\frac{\varsigma}{2}\right)\right]\delta R-2in_{0}\Omega CI_{0}X\Xi.\label{eq:4.22}
\end{equation}
Since the pressure distribution inside the soliton satisfies $\delta p\left(r\right)=\delta p_{\tilde{\vartheta}}\left(r\right)$, the above equation is purely an interface equation.
As the shear force behaves as a second-order smallness near the origin $\varsigma\sim r^{2}$, and decays exponentially at infinity $\varsigma\sim e^{-r}$, which can be obtained from the asymptotic analysis of the stationary equation \eqref{eq:4.1}, the integration of the perturbation equation \eqref{eq:4.22} yields 
\begin{equation}
	\gamma\left(\Omega^{2}+\frac{1}{R^{2}_{0}}- k^{2}\right)\delta R=in_{0}\Omega CI_{0}\left(kR_{0}\right).
\end{equation}
Combining with the condition \eqref{eq:4.19}, one obtains the dispersion relation
\begin{equation}
	\Omega^{2}=\frac{\gamma I_{1}\left(kR_{0}\right)}{\omega n_{0}R^{3}_{0}I_{0}\left(kR_{0}\right)+\gamma R^{2}_{0} I_{1}\left(kR_{0}\right)kR_{0}}\left(k^{2}R^{2}_{0}-1\right)kR_{0}.\label{eq:4.24}
\end{equation}
First, this is a hydrodynamic mode satisfying the definition $\Omega\left(k\rightarrow0\right)=0$.
Second, this mode exhibits dynamical instability in the long-wavelength regime $\lambda>\lambda_{c}$, with a threshold that depends solely on the geometric radius of the Q-string $\lambda_{c}=2\pi R_{0}$.
Finally, on the one hand, in the thin-wall limit the energy density satisfies relation $\epsilon_{0}=\omega n_{0}$; on the other hand, when the surface tension, namely the surface energy density, is much smaller than the bulk energy density of the interior $\gamma\ll\epsilon_{0}$, the second term in the denominator of the above dispersion relation can be neglected.
In this case, the dispersion relation of the membrane instability of Q-strings \eqref{eq:4.24} degenerates to the Rayleigh-Plateau dispersion relation \eqref{eq:2.40}.
Figure \ref{fig:20} shows the numerical results obtained by solving the linearized Klein-Gordon equation without approximation using the spectral decomposition method, demonstrating perfect agreement with the analytical expression \eqref{eq:4.24} in the thin-wall regime \footnote{In our previous work \cite{Chen:2024axd}, we found that the numerical results differed from the Rayleigh-Plateau dispersion relation by a multiplicative factor. This work resolves that issue by recognizing that the density term in the Rayleigh-Plateau dispersion relation refers to the energy density rather than the particle number density. In that work, we mistakenly used the particle number density, which led to a discrepancy by a factor of $\omega^{-1}$.}.
Thus, we have analytically demonstrated that the soliton interface behave like a fluid membrane, subject to the membrane instability induced by surface tension.

\subsection{Surface area and surface energy}
The exponential growth of the membrane instability will break the translational symmetry of the Q-string along the cylindrical direction and drive a topological transition of the interface.
The transition proceeds in the direction that minimizes the surface area, leading the cylindrical surface to split into multiple spheres with a reduced total area.
For a perturbation along the cylindrical direction, with the Q-string length being an integer multiple of the wavelength, $L=N\lambda$ with integer $N$, the radius evolves as
\begin{equation}
	R=R_{0}+\sin\left(kz\right)\delta R\left(t\right)-\frac{1}{4R_{0}}\delta^{2}R\left(t\right)+o\left(\delta^{4}\right),
\end{equation}
where the second-order term arises from the conservation of the particle number
\begin{equation}
	\frac{d}{dt}Q=\pi n_{0}\frac{d}{dt}\int_{0}^{L}R^{2}dz=0.
\end{equation}
In this case, the variation of the surface area of the Q-string is given by
\begin{equation}
	A=2\pi L\left[R_{0}-\frac{1}{4R_{0}}\left(1-R^{2}_{0}k^{2}\right)\delta^{2}R\left(t\right)\right],
\end{equation}
from which it can be observed that perturbations with a wavelength exceeding the threshold of the membrane instability $\lambda_{c}=2\pi R_{0}$ will lead to a reduction in the surface area.
Typically, surface energy scales with surface area, so reducing the latter lowers the former.
In a soliton system, however, this relation breaks down.
From the definition \eqref{eq:4.7}, the variation of the surface energy of the Q-string is given by
\begin{equation}
	E_{\text{surface}}\left(t\right)=2\pi L \gamma\left[R_{0}-\left(1-2R^{2}_{0}k^{2}\right)\frac{\delta^{2}R\left(t\right)}{4R_{0}}\right],
\end{equation}
indicating that the critical wavelength of perturbations that reduce the surface energy is $\tilde{\lambda}_{c}=2\sqrt{2}\pi R_{0}$.
Since this critical wavelength is larger than the threshold of the membrane instability, the initial growth of perturbations exhibits two scenarios depending on the wavelength:
\begin{itemize}
	\item $\tilde{\lambda}_{c}<\lambda$, both the surface area and surface energy decrease;
	\item $\lambda_{c}<\lambda<\tilde{\lambda}_{c}$, the surface area decreases, but the surface energy increases.
\end{itemize}
The above analysis is based on linear perturbation theory; therefore, the results can only indicate that the surface energy may temporarily increase at the initial stage, but they do not determine the behavior of the surface energy in the nonlinear evolution, nor the final state.
The discrepancy between the two critical wavelength arises because the surface area depends on the integral of the modulus of the interface normal vector, while the surface energy is defined through the integral of its square.
At first-order expansion, this leads to a factor-of-two difference for perturbations along the cylindrical direction.
Such a conclusion seems to contradict our empirical experience in nature.
The reason behind this is the assumption that the surface tension of a material is an inherent thermodynamic property, independent of the geometry of the interface, so that the surface energy is proportional to the surface area.
In soliton system, however, the surface tension depends not only on the value of the internal scalar field but also on the geometry of the interface.
For cases with a non-uniform geometric radius $R\left(z\right)$, the surface tension is given by
\begin{equation}
	\tilde{\gamma}=\gamma\sqrt{1+\left(\partial_{z}R\right)^{2}},
\end{equation}
and the surface energy can be expressed as
\begin{equation}
	E_{\text{surface}}=2\pi\int\tilde{\gamma} R\sqrt{1+\left(\partial_{z}R\right)^{2}}dz.
\end{equation}
At this point, if we also neglect the dependence of the surface tension on the geometry along the cylindrical direction, the surface energy would become proportional to the surface area $E_{\text{surface}}=\tilde{\gamma}A$, and the critical wavelength for reducing the surface energy would coincide with the threshold of the membrane instability.
However, this assumption is unreasonable.
Therefore, we can conclude that the membrane instability fundamentally reduces the surface area rather than the surface energy.

\subsection{Physical scenarios}
\begin{figure}
	\centering\includegraphics[width=.8\linewidth]{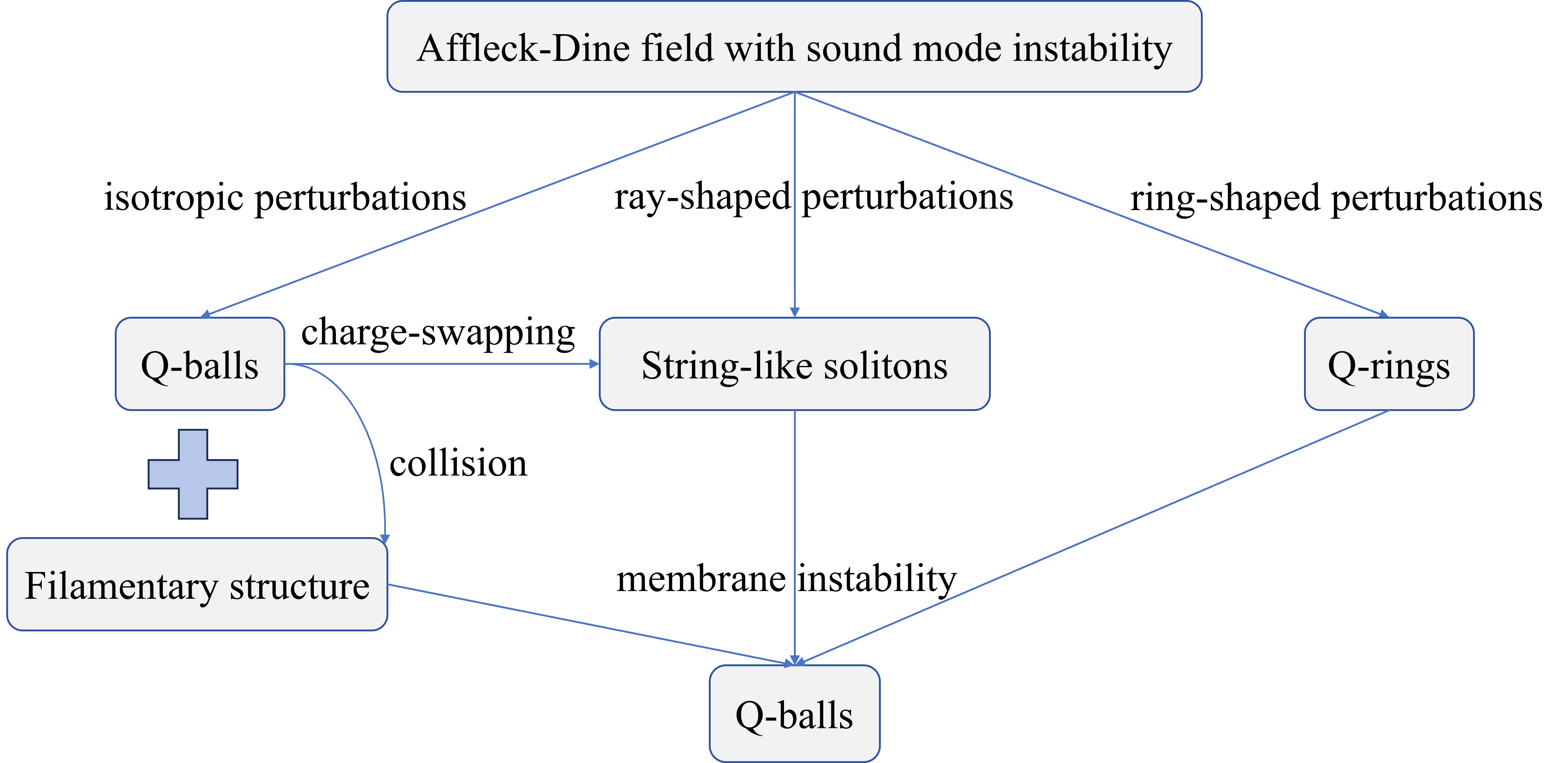}
	\caption{The membrane instability in the formation and dynamics of solitons.}
	\label{fig:21}
\end{figure}

Due to the existence of the wavelength threshold, the onset of the membrane instability for string-like solitons depends on the relative size of their length to the perturbation wavelength, providing two distinct dynamical paradigms:
\begin{itemize}
	\item Threshold wavelength greater than the soliton length ($L<\lambda_{c}$): Since the wavelength of any perturbation is necessarily shorter than the soliton length, the soliton is immune to the membrane instability.
	In this case, the soliton contracts along the cylindrical direction, forming a single large spherical soliton, a Q-ball.
	\item Threshold wavelength smaller than the soliton length ($\lambda_{c}<L$): In this case, perturbations with wavelengths $\lambda_{c}<\lambda<L$ can trigger the membrane instability.
	The soliton interface locally contracts along the cross-sectional direction, gradually leading to interface rupture and eventually forming several smaller spherical Q-balls, analogous to the breakup of a liquid jet.
\end{itemize}
These dynamical paradigms are crucial for the formation and dynamics of solitons.
During the soliton formation process, for random perturbations without directionality, new thermal phases with thermodynamic stability are isotropically generated in the Affleck-Dine field subject to the sound mode instability, resulting in the emergence of spherically symmetric Q-balls.
In this case, nonlinear evolution reveals that the Q-balls are connected by filamentary structures \cite{Enqvist:2000cq,Hiramatsu:2010dx}, which subsequently fragment into smaller Q-balls.
This phenomenon may be attributed to the membrane instability.
In contrast, for directional ray-like perturbations, the soliton phase is expected to condense along the ray direction, forming string-like solitons.
Such configurations may further shrink or fragment, depending on the relative size of their length and cross-sectional circumference.
More intriguingly, ring-shaped perturbations may generate toroidal topological solitons, known as Q-rings.
These solitons have been observed to exhibit two distinct evolution pathways \cite{Axenides:2001pi}, collapse or fragmentation, in agreement with the above predictions.
Similar ring-like Q-matter configurations can also arise during Q-ball collisions \cite{Battye:2000qj}.
In addition, it has been shown that a Q-ball and an anti-Q-ball can remain in equilibrium through charge-swapping behavior \cite{Copeland:2014qra}, thereby allowing the existence of string-like solitons with alternating arrangements.
The stability of such configurations is expected to be governed by the competition between the charge-swapping interaction and the membrane instability, which deserves further investigation.
Since the charge-swapping configuration can be regarded as an oscillon \cite{Alonso-Izquierdo:2025iet}, the membrane properties may thus be naturally generalized to oscillon systems.
These scenarios, summarized in figure \ref{fig:21}, underscore the universality and significance of the membrane instability.

\section{Conclusion and outlook}\label{sec:C}
In this work, we reveal three types of dynamical instability in soliton field theory: the tachyonic instability, the sound mode instability, and the membrane instability.
Their properties are summarized in table \ref{table:2}.
As demonstrated in the text, these three instabilities play important roles in the formation and dynamics of solitons, especially the latter two hydrodynamic instabilities.

\begin{table}[h!]
	\small
	\centering
	\begin{tabular}{c|c|c|c}
		\hline
		&Tachyonic instability& Sound mode instability& Membrane instability\\
		\hline
		\hline
		\multirow{3}{*}{Basic features}&non-hydrodynamic&\multicolumn{2}{c}{hydrodynamic}\\
		\cline{2-4}
		&maintain spatial symmetry&\multicolumn{2}{c}{break spatial symmetry}\\
		\cline{2-4}
		&non-directional&non-directional&directional\\
		\hline
		Mechanism&negative mass square&thermodynamic instability&surface tension\\
		\hline
		Threshold&-&viscosity&geometric radius\\
		\hline
		Transition&excited to ground state&thermodynamic properties&surface topology\\
		\hline
		Examples&spontaneous scalarization&holographic fluid&fluid jet\\
		\hline
	\end{tabular}
	\caption{Properties of dynamical instabilities in soliton field theory.}
	\label{table:2}
\end{table}

On the one hand, we demonstrate that the essence of the soliton formation mechanism is the sound mode instability induced by thermodynamic instability.
For a uniform soliton fluid, there exists a non-hydrodynamic mode and a hydrodynamic mode in the long-wavelength regime.
The dispersion relation of the non-hydrodynamic mode resembles that of a real scalar field, exhibiting tachyonic instability in the case of negative effective mass squared.
The hydrodynamic mode is similar to the sound mode in fluids, which exhibits undamped propagation in the thermodynamically stable case, resembling an ideal fluid, and exhibits the sound mode instability with a threshold in the thermodynamically unstable case, resembling a viscous fluid.
In soliton field theory, there exist two types of soliton fluids: one located at the local maximum of the effective potential, corresponding to an excited state, and the other is a local ground state located at a local minimum.
As expected, the ground state is dynamically stable, while the excited state exhibits dynamical instability.
Interestingly, such an instability can be either tachyonic or sound mode instability, depending on the type of scalar potential and the parameter regime.
Different types of instabilities lead to different evolution pathways.
An excited state subject to the sound mode instability undergoes phase separation, in which new thermal phases with thermodynamic stability are generated to produce solitons.
These solitons can be regarded as a coexistence state, in which an internal matter phase with positive pressure and an external vacuum phase achieve dynamical equilibrium through an interface that provides surface tension.
From the perspective of phase transition, the existence of vacuum bubbles, characterized by an internal vacuum phase surrounded by an external matter phase with negative pressure, is naturally predicted.
An excited state with tachyonic instability admits two possible evolution routes: one is decay into a ground-state soliton fluid, and the other is evolution into an excited state with sound mode instability, which serves as a dynamical intermediate state and subsequently undergoes phase separation into solitons or vacuum bubbles.

On the other hand, we demonstrate that the soliton interface resembles a fluid membrane, subject to the membrane instability induced by surface tension.
Indications of this phenomenon can be seen from the stationary equation of the soliton system, which exhibits the structure of the Young-Laplace equation, describing the pressure generated by the surface tension and curvature of the interface. 
In the thin-wall limit, and assuming the interface has zero thickness and the interior matter is incompressible, the dispersion relation for the membrane instability is analytically derived, which degenerates to the Rayleigh-Plateau dispersion relation when the surface energy density is smaller than the bulk energy density.
Due to the existence of the wavelength threshold for this membrane instability, for solitons with local string structures, there are two distinct evolution paradigms, depending on the relative size of their local length compared to the threshold.
For situations where the local length is less than the instability threshold, such structures are immune to the membrane instability but may contract along the axial direction.
In the opposite case, when the local length exceeds the threshold, the membrane instability can trigger a topological transition of the interface, splitting it into several spheres with a smaller total surface area.
In this process, we find that for the wavelength of perturbations, there is a region near the instability threshold, where the perturbation can increase the surface energy of the system in the initial stage of the evolution.
Such results indicate that the essence of the membrane instability is the reduction of surface area rather than of surface energy.
This membrane instability has been found to be widely present in physical scenarios such as soliton formation and dynamics.

The hydrodynamic properties in soliton field theory reveal a duality between solitons and fluids.
This duality is not only essential for investigating the equilibrium and non-equilibrium properties of solitons, but also provides insights into various physical phenomena in cosmology and gravity.
In cosmological research, phenomena such as the cosmological phase transitions \cite{Hindmarsh:2013xza} and the formation of primordial black holes \cite{Niemeyer:1999ak} typically rely on the description of hydrodynamic models.
In these processes, the generation and propagation of sound waves often lead to the further excitation of gravitational waves.
Given that soliton fluids also possess the sound mode, they provide a promising candidate framework as an alternative to hydrodynamic models in cosmology.
Moreover, the expansion of the early universe can be described as the growth of the vacuum phase within a fluid with negative pressure \cite{Guth:1980zm}.
Similarly, in certain cosmological models, dark energy is also considered as a fluid with negative pressure \cite{Kamenshchik:2001cp,Bento:2002ps}.
In soliton field theory, there exist two types of soliton fluid with negative pressure: one exhibits sound mode instability, while the other is dynamically stable.
The expansion of the early universe can be modeled by the generation of vacuum bubbles through phase separation within a soliton fluid with sound mode instability.
The dynamically stable one, on the other hand, could serve as a candidate for dark energy.
Under this assumption, the final state of phase separation depicts a scenario in which the vacuum is surrounded by dark energy.
These scenarios highlight the potential applications of soliton fluids with negative pressure.
Moreover, owing to the intricate connections between oscillons and solitons \cite{Blaschke:2024dlt,Blaschke:2025anm}, such important cosmological objects may naturally be incorporated into the hydrodynamic framework introduced in this work, which deserves further exploration.

For gravitational systems, the horizon of a black hole can be intuitively considered as a fluid membrane \cite{Thorne:1986iy}, with surface gravity playing the role of surface tension.
Furthermore, higher-dimension black strings have been shown to be dynamically unstable to long-wavelength perturbations \cite{Gregory:1993vy}, with a threshold associated with the geometric radius \cite{Cardoso:2006ks}.
The nonlinear evolution indicates that the dynamical behavior resembles that of a low-viscosity jet \cite{Lehner:2010pn}, highlighting the hydrodynamic nature of the horizon of black objects.
Differently, due to black hole thermodynamics, there are indications that this type of instability of black strings may be jointly governed by the membrane instability and the sound mode instability \cite{Emparan:2009cs,Emparan:2009at}.
Accordingly, the hydrodynamic properties of the soliton interface on the one hand provide a guarantee for simulating black objects, and on the other hand suggest that interfaces of matter-composed compact objects may exhibit similar hydrodynamic behavior, as confirmed on the boson string \cite{Herdeiro:2025lwf}, a gravitational derivative of the Q-string.

In addition, this duality provides a field theory tool for studying hydrodynamics.
Soliton fluids exhibiting sound mode instability can be used to investigate the dynamical behavior of fluids during phase separation, such as the transport behavior of physical quantities and the formation and dynamics of domain walls.
Soliton states can serve as models for liquids with free surfaces, revealing the real-time dynamics of liquid configurations such as shallow water, jet, droplets, and structures with other topology.
Another point we wish to emphasize is that hydrodynamic instabilities are universal and may be widely present in various physical systems, serving as a typical mechanism for breaking spatial symmetries.


\appendix
\section{Fluid dynamics in the Landau-Lifshitz frame}\label{sec:Aa}
For completeness, this appendix presents the fluid dynamic formulation in the Landau-Lifshitz frame.
In contrast to the explicit definition adopted in the Eckart frame, the fluid four-velocity in the Landau-Lifshitz frame is implicitly defined by
\begin{equation}
	u^{\mu}=u_{\nu}T^{\mu\nu}/\sqrt{T^{\alpha\nu}T^{\beta}_{\nu}u_{\alpha}u_{\beta}}.
\end{equation}
In this case, the dissipative quantities are $\{\Pi,n^{\mu},\pi^{\mu\nu}\}$ without the energy diffusion current $q^{\mu}=0$.
Similarly, the constitutive relations can be constructed from the second law of thermodynamics.
The procedure follows the same steps as presented in subsection \ref{sec:Si}.
Expanding the dissipative correction of the entropy four-current up to second order
\begin{equation}
	TF^{\gamma}=\left(\alpha_{1}\Pi+\alpha_{\Pi}\Pi^{2}-\alpha_{n}n_{\mu}n^{\mu}+\alpha_{\pi}\pi_{\mu\nu}\pi^{\mu\nu}\right)u^{\gamma}+\beta_{1} n^{\gamma}+\beta_{\Pi}\Pi n^{\gamma}+\beta_{\pi}\pi^{\mu\gamma}n_{\mu},
\end{equation}
with expansion coefficients $\{\alpha_{i},\beta_{i}\}$, the entropy production rate takes the following form
\begin{equation}
	\partial_{\mu}S^{\mu}=T^{-1}\alpha_{1}D\Pi+T^{-1}\left(\beta_{1}+\mu\right)\nabla_{\mu}n^{\mu}+\Pi f_{\Pi}-n_{\mu}f^{\mu}_{n}+\pi_{\mu\nu}f^{\mu\nu}_{\pi}.
\end{equation}
The requirement of semi-positivity imposes the conditions $\alpha_{1}=0$, $\beta_{1}=-\mu$ and 
\begin{equation}
	f_{\Pi}=\frac{\Pi}{c_{\Pi} T},\quad f^{\mu}_{n}=\frac{n^{\mu}}{c_{n} T},\quad f^{\mu\nu}_{\pi}=\frac{\pi^{\mu\nu}}{c_{\pi}T},\label{eq:A4}
\end{equation}
with positive proportionality coefficients $\{c_{i}\}$.
Defining the relaxation times
\begin{equation}
	\tau_{\Pi}=-2c_{\Pi}\alpha_{\Pi},\quad \tau_{n}=-2c_{n}\alpha_{n},\quad \tau_{\pi}=-2c_{\pi}\alpha_{\pi},
\end{equation}
the relations \eqref{eq:A4} yield the following relaxation-type constitutive equations for the dissipative quantities
\begin{subequations}
	\begin{align}
		\tau_{\Pi}D\Pi+\Pi&=-c_{\Pi}\theta+\delta_{\Pi},\\
		\tau_{n}\Delta^{\mu}_{\nu}Dn^{\nu}+n^{\mu}&=c_{n}T\nabla^{\mu}\left(\frac{\mu}{T}\right)+\delta^{\mu}_{n},\\
		\tau_{\pi}\Delta_{\alpha}^{\mu}\Delta_{\beta}^{\nu}D\pi^{\alpha\beta}+\pi^{\mu\nu}&=c_{\pi}\sigma^{\mu\nu}+\delta^{\mu\nu}_{\pi},
	\end{align}\label{eq:A6}
\end{subequations}
with the second-order corrections
\begin{subequations}
	\begin{align}
		\delta_{\Pi}&=-\frac{\tau_{\Pi}}{2}\left[\theta+D\ln\left(\frac{\tau_{\Pi}}{c_{\Pi}T}\right)\right]\Pi+c_{\Pi}\beta_{\Pi}\left[\nabla-c_{1}a+c_{2}\nabla\ln\left(\beta_{\Pi}/T\right)\right]_{\mu}n^{\mu},\\
		\delta^{\mu}_{n}&=-\frac{\tau_{n}}{2}\left[\theta+D\ln\left(\frac{\tau_{n}}{c_{n}T}\right)\right]n^{\mu}-c_{n}\beta_{\Pi}\left[\nabla-\left(1-c_{1}\right)a+\left(1-c_{2}\right)\nabla\ln\left(\beta_{\Pi}/T\right)\right]^{\mu}\Pi\nonumber\\
		&\quad-c_{n}\beta_{\pi}\Delta^{\mu}_{\alpha}\left[\nabla-\left(1-c_{3}\right)a+\left(1-c_{4}\right)\nabla\ln\left(\beta_{\pi}/T\right)\right]_{\nu}\pi^{\alpha\nu},\\
		\delta^{\mu\nu}_{\pi}&=-\frac{\tau_{\pi}}{2}\left[\theta+D\ln\left(\frac{\tau_{\pi}}{c_{\pi}T}\right)\right]\pi^{\mu\nu}+c_{\pi}\beta_{\pi}\left[\Delta^{\mu}_{\alpha}\Delta^{\nu}_{\beta}\left[\nabla-c_{3}a+c_{4}\nabla\ln\left(\beta_{\pi}/T\right)\right]^{(\alpha}n^{\beta)}\right.\nonumber\\
		&\left.\quad-\left[\nabla-c_{3}a+c_{4}\nabla\ln\left(\beta_{\pi}/T\right)\right]_{\alpha}n^{\alpha}\Delta^{\mu\nu}/3\right].
	\end{align}
\end{subequations}
Such results indicate that the particle diffusion current in the Landau-Lifshitz frame is equivalent to the energy diffusion current in the Eckart frame.

As expected, the dispersion relations in the two frames exhibit equivalent structures.
Substituting the perturbations
\begin{equation}
	\begin{aligned}
		n&=n_{0}+\delta_{x} n,\quad& \epsilon&=\epsilon_{0}+\delta_{x} \epsilon, \quad& u^{\mu}&=\left(1,\vec{0}\right)+\left(0,\delta_{x} u^{i}\right),\\
		\Pi&=\delta_{x} \Pi,\quad& n^{\mu}&=\left(0,\delta_{x}n^{i}\right), \quad& \pi^{\mu\nu}&=\left[\left(0,\vec{0}\right),\left(\vec{0},\delta_{x}\pi^{ij}\right)\right],
	\end{aligned}
\end{equation}
into the fluid dynamic equations \eqref{eq:2.7} and the constitutive equations \eqref{eq:A6}, one obtains three types of dispersion relations:

\noindent the longitudinal dispersion relation
\begin{equation}
	\begin{aligned}
		&0=\left[\Omega-\left(h^{-1}_{0}\frac{\partial p}{\partial n}+\frac{\partial p}{\partial\epsilon}\right)\frac{k^{2}}{\Omega}-\frac{K_{\Pi}+K_{\pi}}{\epsilon_{0}+p_{0}}\right]\left[c^{-1}_{n}\left(\tau_{n}\Omega+i\right)-T\frac{\partial\tilde{\mu}}{\partial n}\frac{k^{2}}{\Omega}-\beta^{2}_{\Pi}K_{\Pi}-\beta^{2}_{\pi}K_{\pi}\right]\\
		&-\left(\beta_{\Pi}K_{\Pi}-\beta_{\pi}K_{\pi}-\frac{\partial p}{\partial n}\frac{k^{2}}{\Omega}\right)\left[\frac{\beta_{\Pi}K_{\Pi}-\beta_{\pi}K_{\pi}}{\epsilon_{0}+p_{0}}-T\left(h^{-1}_{0}\frac{\partial\tilde{\mu}}{\partial n}+\frac{\partial\tilde{\mu}}{\partial\epsilon}\right)\frac{k^{2}}{\Omega}\right],
	\end{aligned}
\end{equation}

\noindent the transverse dispersion relation
\begin{equation}
	0=\left[\left(\epsilon_{0}+p_{0}\right)\Omega-\frac{3}{4}K_{\pi}\right]\left[c^{-1}_{n}\left(\tau_{n}\Omega+i\right)-\frac{3}{4}\beta^{2}_{\pi}K_{\pi}\right]-\frac{9}{16}\beta^{2}_{\pi}K^{2}_{\pi},
\end{equation}

\noindent and the damped dispersion relation
\begin{equation}
	0=\tau_{\pi}\Omega+i,
\end{equation}
with the thermodynamic potential $\tilde{\mu}=\mu/T$.
In the Eckart frame, the transverse dispersion relation develops a dynamical instability in the Navier-Stokes limit \eqref{eq:2.24}, revealing the pathological nature of the relativistic Navier-Stokes theory.
In the Landau-Lifshitz frame, such instability seems absent at first glance
\begin{subequations}
	\begin{align}
		\Omega_{\pi}&=-\frac{1}{2}\frac{c_{\pi}i}{\epsilon_{0}+p_{0}}k^{2}+o(k^{4}),\\
		\Omega_{d}&=-\tau^{-1}_{n}i+o(k^{2}).
	\end{align}
\end{subequations}
However, this result does not imply that the relativistic Navier-Stokes theory is self-consistent in the Landau-Lifshitz frame, since the instability emerges in the dispersion relation after a Lorentz transformation \cite{Hiscock:1985zz}.
The longitudinal dispersion relation gives rise to the particle diffusion mode and the sound mode
\begin{equation}
	\begin{aligned}
		\Omega_{n}&=-\frac{c_{n}h^{2}i}{nTc_{\hat{s}}}k^{2}+o(k^{4}),\\
		\Omega_{s}&=v_{s}k-i\Gamma k^{2}+o(k^{3}),
	\end{aligned}
\end{equation}
with the sound attenuation coefficient
\begin{equation}
	\Gamma=\frac{1}{2\left(\epsilon_{0}+p_{0}\right)}\left[c_{\Pi}+\frac{2}{3}c_{\pi}+\left(\frac{v_{s}c_{\epsilon}}{nc_{\hat{s}}}\right)^{2}c_{n}h^{2}\right]i.
\end{equation}
from which it can be observed that the two frames are identified through relation $c_{q}=c_{n}h^{2}$.

\section{Thermodynamic relations}\label{sec:Tr}
This appendix provides the thermodynamic relations required for deriving the dispersion relations within the framework of fluid dynamics.
Since only two independent thermodynamic variables are involved, all other thermodynamic quantities can be expressed as functions of these two.
However, the arbitrariness in the choice of independent variables gives rise to a series of corresponding thermodynamic relations.
Starting from the first law of thermodynamics
\begin{subequations}
	\begin{align}
		d\epsilon&=nTd\hat{s}+hdn,\\
		dp&=nhd\ln T+nTd\tilde{\mu},
	\end{align}
\end{subequations}
one can obtain the following six thermodynamic derivatives
\begin{subequations}
	\begin{align}
		&\left[\frac{\partial\epsilon}{\partial\hat{s}}\right]_{n}=nT,&\quad &\left[\frac{\partial\epsilon}{\partial n}\right]_{\hat{s}}=h,&\quad &\left[\frac{\partial\hat{s}}{\partial n}\right]_{\epsilon}=-\frac{h}{nT},\\
		&\left[\frac{\partial p}{\partial T}\right]_{\tilde{\mu}}=\frac{nh}{T},&\quad &\left[\frac{\partial p}{\partial\tilde{\mu}}\right]_{T}=nT,&\quad &\left[\frac{\partial T}{\partial\tilde{\mu}}\right]_{p}=-\frac{T^{2}}{h}.
	\end{align}
\end{subequations}
Furthermore, the integrability conditions yield the following Maxwell relations
\begin{subequations}
	\begin{align}
		&\left[\frac{\partial p}{\partial\hat{s}}\right]_{n}=n^{2}\left[\frac{\partial T}{\partial n}\right]_{\hat{s}},&\quad &\left[\frac{\partial T}{\partial n}\right]_{\epsilon}=T^{2}\left[\frac{\partial\tilde{\mu}}{\partial\epsilon}\right]_{n},&\quad&\left[\frac{\partial p}{\partial\hat{s}}\right]_{\epsilon}=-n^{2}T^{2}\left[\frac{\partial\tilde{\mu}}{\partial \epsilon}\right]_{\hat{s}},\\
		&\left[\frac{\partial n}{\partial T}\right]_{p}=n^{2}\left[\frac{\partial\hat{s}}{\partial p}\right]_{T},&\quad &\left[\frac{\partial\epsilon}{\partial\tilde{\mu}}\right]_{T}=T^{2}\left[\frac{\partial n}{\partial T}\right]_{\tilde{\mu}},&\quad&\left[\frac{\partial\epsilon}{\partial \tilde{\mu}}\right]_{p}=-n^{2}T^{2}\left[\frac{\partial\hat{s}}{\partial p}\right]_{\tilde{\mu}}.
	\end{align}
\end{subequations}
On this basis, all other thermodynamic relations can be derived from the above formulas combined with the chain rule
\begin{subequations}
	\begin{align}
		\left[\frac{\partial A}{\partial B}\right]_{C}&=\left[\frac{\partial A}{\partial D}\right]_{C}\left[\frac{\partial D}{\partial B}\right]_{C},\\
		\left[\frac{\partial A}{\partial B}\right]_{C}&=\left[\frac{\partial A}{\partial B}\right]_{D}+\left[\frac{\partial A}{\partial D}\right]_{B}\left[\frac{\partial D}{\partial B}\right]_{C}.
	\end{align}
\end{subequations}


\section*{Acknowledgement}
The author acknowledges the support from the National Natural Science Foundation of China with Grant No. 12447129 and the fellowship from the China Postdoctoral Science Foundation with Grant No. 2024M760691.

\bibliography{references}

\end{document}